\documentstyle[prb,aps]{revtex}
\begin{document}
\draft


\title{An Efficient Molecular Dynamics Scheme for the Calculation of
Dopant Profiles due to Ion Implantation}

\author{Keith M. Beardmore and
Niels Gr{\o}nbech-Jensen}

\address{Theoretical Division,
Los Alamos National Laboratory, Los Alamos, NM 87545, USA}

\date{\today}
\maketitle

\begin{abstract}
   We present a highly efficient molecular dynamics scheme for calculating
the concentration depth profile of dopants in ion irradiated
materials. The scheme incorporates several methods for reducing the
computational overhead, plus a rare event algorithm that allows
statistically reliable results to be obtained over a range of
several orders of magnitude in the dopant concentration.
   We give examples of using this scheme for calculating concentration
profiles of dopants in crystalline silicon. Here we can predict the
experimental profile over five orders of magnitude
for both channeling and non-channeling implants at energies up to 100s
of keV.
   The scheme has advantages over binary collision approximation (BCA)
simulations, in that it does not rely on a large set of empirically
fitted parameters. Although our scheme has a greater computational
overhead than the BCA, it is far superior in the low ion energy regime,
where the BCA scheme becomes invalid.
\end{abstract}

\pacs{PACS numbers: 34.10.+x, 82.20.Wt, 61.72.Tt, 82.80.Ms}


\renewcommand{\thefootnote}{\fnsymbol{footnote}}

\section{Introduction}
The principal reason for implanting ions\footnote{Within
this paper, \emph{ion} is used to refer to the
implanted species, and \emph{atom} to refer to a
particle of the target material;
this has no implication to the charge state of either atom type.}
into silicon wafers is to dope
regions within the substrate, and hence modify their electrical properties
in order to create electronic devices.
The quest for ever increasing processor performance demands
smaller device sizes.
The measurement and modeling of dopant profiles within these
ultra shallow junction devices is challenging, as effects that were negligible
at high implant energies become increasingly important as the implant energy
is lowered.
The experimental measurement of dopant profiles by secondary ion mass
spectrometry (SIMS) becomes problematic for very low energy
(less than 10 keV) implants.
There is a limited depth resolution of measured profiles
due to profile broadening,
as the SIMS ion-beam produces `knock-on's,
and so leads to effects such as diffusion of dopants and mixing.
The roughness and disorder of the sample surface can also convolute
the profile, although this can be avoided to a large extent by
careful sample preparation\cite{par96}.

The use of computer simulation as a method for studying the effects of
ion bombardment of solids is well established.
Binary collision approximation\cite{rob74} (BCA),
`event-driven' codes have traditionally been used to
calculate such properties
as ranges of implanted species and the damage distributions resulting
from the collision cascade.
In this model, each ion trajectory is constructed as a series of repulsive
two-body encounters with initially stationary target atoms, and with
straight line motion between collisions.
Hence the algorithm consists of finding the next collision partner,
and then calculating the asymptotic motion of the ion after the collision.
This allows for efficient
simulation, but leads to failure of the method at low ion energies.
The BCA approach breaks down when
multiple collisions (where the ion has simultaneous interactions
with more than one target atom) or collisions between moving atoms
become significant,
when the crystal binding energy is of the same order as the energy
of the ion,
or when the time spent within a collision is too long for the calculation
of asymptotic trajectories to be valid.
Such problems are clearly evident when one attempts to use the BCA to
simulate channeling in semiconductors; here the interactions between
the ion and the target are neither binary nor collisional in nature,
rather they occur as many simultaneous soft interactions which steer
the ion down the channel. 

An alternative to the BCA is to use molecular dynamics (MD)
simulation, which has long been applied to the investigation of ion
bombardment of materials\cite{vin60,har88},
to calculate the ion trajectories\cite{bea97,cai96}.
The usefulness of this approach was once limited by its computational cost
and the lack of realistic models to describe materials.
With the increase in computational power, the development of
efficient algorithms, and the production of accurate empirical potentials,
it is now feasible to conduct realistic MD simulations.
In the classical MD model, atoms are represented by point
masses that interact via an empirical potential function that is typically
a function of bond lengths and angles; in the case of Si a three-body or
many-body potential, rather than a pair potential is required to model
the stable diamond lattice and to account for the bulk crystal
properties.
The trajectories of atoms are
obtained by numerical integration of Newton's laws, where
the forces are obtained from the analytical derivative of the potential
function.
Thus, MD provides a far more realistic description of the
collision processes than the BCA,
but at the expense of a greater computational requirement.
   Here we present a highly efficient MD scheme that is optimized
to calculate the concentration profiles of ions
implanted into crystalline silicon.
The algorithms are incorporated into our implant modeling
molecular dynamics code, REED\footnote{Named for
`Rare Event Enhanced Domain following'
molecular dynamics.},
which runs on many architectures either as a serial,
or as a trivially parallel program.

\section{Molecular Dynamics Model}
   The basis of the molecular dynamics model is a collection of empirical
potential functions that describe interactions between atoms and give
rise to forces between them. In addition to the classical interactions
described by the potential functions, the interaction of the ion with the
electrons within the target is required for ion implant simulations,
as this is the principle way in
which the ion loses energy. This is accomplished via a
phenomenological electronic stopping-power model.
Other ingredients necessary to the computation are a
description of the target material structure and thermal vibration
within the solid.
It is also necessary to define a criterion to decide when the ion
has come to rest in the substrate. We terminate a trajectory
when the \emph{total} energy of the ion
falls below 5 eV. This was chosen to be well below the displacement
threshold energy of Si (around 20 eV)\cite{mil94}.

\subsection{Empirical Potential Functions}
   Interactions between Si atoms are modeled by a many-body potential
developed by Tersoff\cite{ter88}. This consists of Morse-like repulsive
and attractive pair functions of interatomic separation, where the attractive
component is modified by a many-body function that has the role of
an effective Pauling bond order.
The many-body term incorporates information about the local environment of
a bond;
due to this formalism the potential can describe features
such as defects and surfaces, which are
very different to the tetrahedral diamond structure.

   ZBL `pair specific' screened Coulomb potentials\cite{zei85} are used
to model the ion-Si interactions for As, B, and P ions.
Where no `pair specific' potential was available, the ZBL `universal'
potential has been used. This is smoothly truncated
with a cosine cutoff between 107\% and 147\% of the sum of the
covalent radii of the atoms involved; the cutoff distances were chosen as
they give
a screening function that approximates the `pair specific' potentials for the
examples available to us.

The ZBL `universal' potential is also used to describe
the close-range repulsive part of the Tersoff Si-Si potential, as the
standard form is not sufficiently strong for small atomic separations.
The repulsive Morse term is splined to a shifted ZBL potential,
by joining the two functions at the point where they are co-tangent.
In the case of Si-Si interactions, the join is at
an atomic separation of 0.69 \AA, and requires the ZBL function to be
shifted by
148.7 eV. The increase in the value of the short-range repulsive
potential compensates for the attractive part of the Tersoff potential,
which is present even at short-range.

\subsection{Inelastic Energy Loss}
The Firsov model\cite{fir59} is used to describe the loss of kinetic
energy from the ion due to inelastic collisions with target atoms.
We implement this using a velocity dependent pair potential, as derived
by Kishinevskii\cite{kis62}.
This gives the force between atoms $i$ and $j$ as:
\begin{equation}
{\bf F}_{ij} = {2^{1/3} \hbar \over 2 \pi a_{\text{B}}}({\bf v}_j -{\bf v}_i)
\Biggl( Z^2_1 \times I\biggl(
{Z_1^{1/3} \alpha R \over a}\biggr) + 
Z^2_2 \times I\biggl(
{Z_2^{1/3} (1-\alpha) R \over a}\biggr) \Biggr)
\label{firs1}
\end{equation}
where:
\begin{eqnarray}
I(X) = \int^{\infty}_{X} {\chi^2(x) \over x} dx \text{, and }
\alpha = \Biggl(1+\biggl({Z_2 \over Z_1} \biggr)^{1/6}\Biggr)^{-1}
\label{firs2}
\end{eqnarray}
and $\chi(x)$ is a screening function, $Z$ is atomic number ($Z_1 > Z_2$),
$R$ is atomic separation, and $a =(9\pi^2/128)^{1/3} a_{\text{B}}$.
For consistency with the ion-Si interactions, we use the
ZBL `universal' screening function within the integral;
there are no fitted parameters in this model.


We have found that it is necessary to include energy loss
due to inelastic collisions, and energy loss due to
electronic stopping (described below) as two distinct mechanisms.
It is not possible to assume that one, or other, of these processes is
dominant and \emph{fit} it to model all energy loss for varying energies and
directions.

\subsection{Electronic Stopping Model}
A new model that involves both global and local contributions to the
electronic stopping is used for the electronic energy
loss\cite{cai96,cai97,cai97a}.
This modified Brandt-Kitagawa\cite{bra82} model was developed for
semi-conductors and contains only one fitted parameter per ion species,
for all energies and incident directions.
We believe that by using a realistic stopping model,
with the minimum of fitted parameters,
we obtain a greater transferability to the modeling of implants
outside the fitting set.
This should be contrasted to many BCA models which require
completely different models for different ion species or even for different
implant angles for the same ion species, and that contain several fitted
parameters per species\cite{hob96,pos92}.

Our model has been successfully used to describe the implant of
As, B, P, and Al ions with energies in the sub MeV range
into crystalline Si in the
$\langle$100$\rangle$, $\langle$110$\rangle$, and
non-channeling directions, and also into amorphous Si.
       While initially developed for use in BCA simulations,
the only modification required to the model for its use in MD is to allow
for the superposition of overlapping charge distributions,
due to the fact that the ion is usually interacting with more than one
atom at a time.
The one fitting parameter is $r^0_s$, the `average' one electron
radius of the target material,
which is adjusted to account for oscillations in the $Z_1$ dependence
of the electronic stopping cross-section\cite{smi97}.

\subsection{Structure of the Target Material}
   For the calculations presented here, the target is crystalline Si
with a surface amorphous layer.
The amorphous structure was obtained from a simulation
of repeated radiation damage and annealing, of an initially
crystalline section of material\cite{bea97a}.
   Thermal vibrations of atoms are modeled by displacing atoms from their
lattice sites using a Debye model. We use a Debye temperature of 519.0 K
for Si obtained by recent electron channeling measurements\cite{bus97}.
This gives an rms thermal vibrational amplitude in one dimension
of 0.0790 \AA\ at 300.0 K.
Note, we do not use the Debye temperature as a fitting parameter
in our model, as is often done in BCA models\cite{hob96}.
The thermal velocity of the atoms is unimportant as it is
so small compared to the ion velocity, and is set to zero.

  At present there is no accumulation of damage within our simulations,
as we wish to verify the fundamental model with the absolute minimum
of parameters that can be fit.
At a later date we will incorporate a
statistical damage model into our simulations in a manner similar to
that used in BCA codes.
   We also intend to include the capability of using amorphous, or
polycrystalline targets in our simulations.

\section{Efficient Molecular Dynamics Algorithms}
During the time that MD has been in use,
many algorithms have been developed
to enhance the efficiency of simulations.
Here we apply a combination of methods to increase the efficiency of
the type of simulation that we are interested in.
We incorporate both widely used methods,
which are briefly mentioned below, and new or lesser known
algorithms for this specific type of simulation
which we describe in greater detail.

\subsection{Basic Algorithms}
   We employ neighbor lists\cite{bea95,ver67,hoc81}
to make the potential and force
calculation O($N$), where $N$ is the number of particles.
Coarse grained cells are used in the construction of the neighbor list;
this is combined with a
Verlet neighbor list algorithm to minimize the size of the list.
Atoms within 125\% of the largest interaction distance are stored in the
neighbor list, which is updated only when the relative motion of atoms is
sufficient for interacting neighbors to have changed.

\subsection{Timestep selection}
  The paths of the atoms are integrated using Verlet's
algorithm\cite{ver67}, with a variable timestep that is dependent upon both
kinetic and potential energy of atoms\cite{bea95}.
For high energy simulations the potential energy as well as the velocity
of atoms is important, as atoms may be moving slowly but have high,
and rapidly changing,
potential energies during impacts. The timestep is selected using:
\begin{equation}
\Delta t_{n} = {C_{DIS} \over
\sqrt{\begin{array}{c}\max \\ \scriptstyle 1 \le i \le N \end{array}
\biggl({\frac{\displaystyle 2 \times [KE_{i} + \max (0, PE_{i})]}
{\displaystyle M_{i}}}\biggr)
}}
\label{tstep}
\end{equation}
where $KE_i$, $PE_i$ and $M_i$ are the kinetic energy, potential energy and
mass respectively of atom $i$, and $C_{DIS}$ is a constant with a value of
0.10 \AA. Away from hard collisions, only the kinetic energy term is
important, and the timestep is selected to give the fastest atom
a movement of $C_{DIS}$ in a single timestep.
When the timestep is increasing, it is limited by:
\begin{equation}
\Delta t_{n}' = \min (1.05 \times \Delta t_{n-1} ,
\case{3}{4} \Delta t_{n-1} + \case{1}{4} \Delta t_{n} )
\label{tincr}
\end{equation}
to prevent rapid oscillations in the size of the timestep, and
the maximum timestep is limited to 2.0 fs.

The timestep selection scheme
was checked to ensure that the total energy in a full (i.e., without the
modifications described below) MD simulation was well
conserved for any single ion implant
with no electronic stopping;
e.g. in the case of a non-channeling (10$^{\circ}$ tilt
and 22$^{\circ}$ rotation) 5 keV
As ion into  a 21168 atom Si\{100\} target,
the energy change was 3.6 eV (0.004\%) during
the 250 fs it took the ion to come to rest.

\subsection{Domain following}
Even with the computation resources available today it is infeasible to
calculate dopant profiles by full MD simulation. Although the method is
O($N$) in the number of atoms involved, the computational requirements
scale extremely quickly with the ion energy.
The cost of the simulation can be estimated as the number of atoms
in the system multiplied by the number of timesteps required. 
Consider the case of an ion, subject to an
energy loss proportional to its velocity, $v(t)$, which is then given by
$v(t) = u \exp(-\alpha t)$ where $u$ is its initial velocity and $\alpha$
is the loss coefficient.
Each dimension of the system must scale approximately as the initial ion
velocity, $u$, to fully contain an ion-path.
If the timestep size is chosen so that the maximum distance moved
by any particle in a single step is constant, the number of timesteps is
approximately proportional to the ion distance.

Hence the method is roughly O($u^{4}$).
Although it is possible to compute a few trajectories at
ion energies of up to 100s of keV, the calculation of the thousands necessary
to produce statistically reliable dopant profiles is out of the question.
Therefore, we have concentrated on developing a restricted MD scheme
which is capable of producing accurate dopant profiles with a much
smaller computational overhead.

\begin{figure}
\vbox{\vspace{-1.20 in} \hspace{0.6 in}
\includegraphics{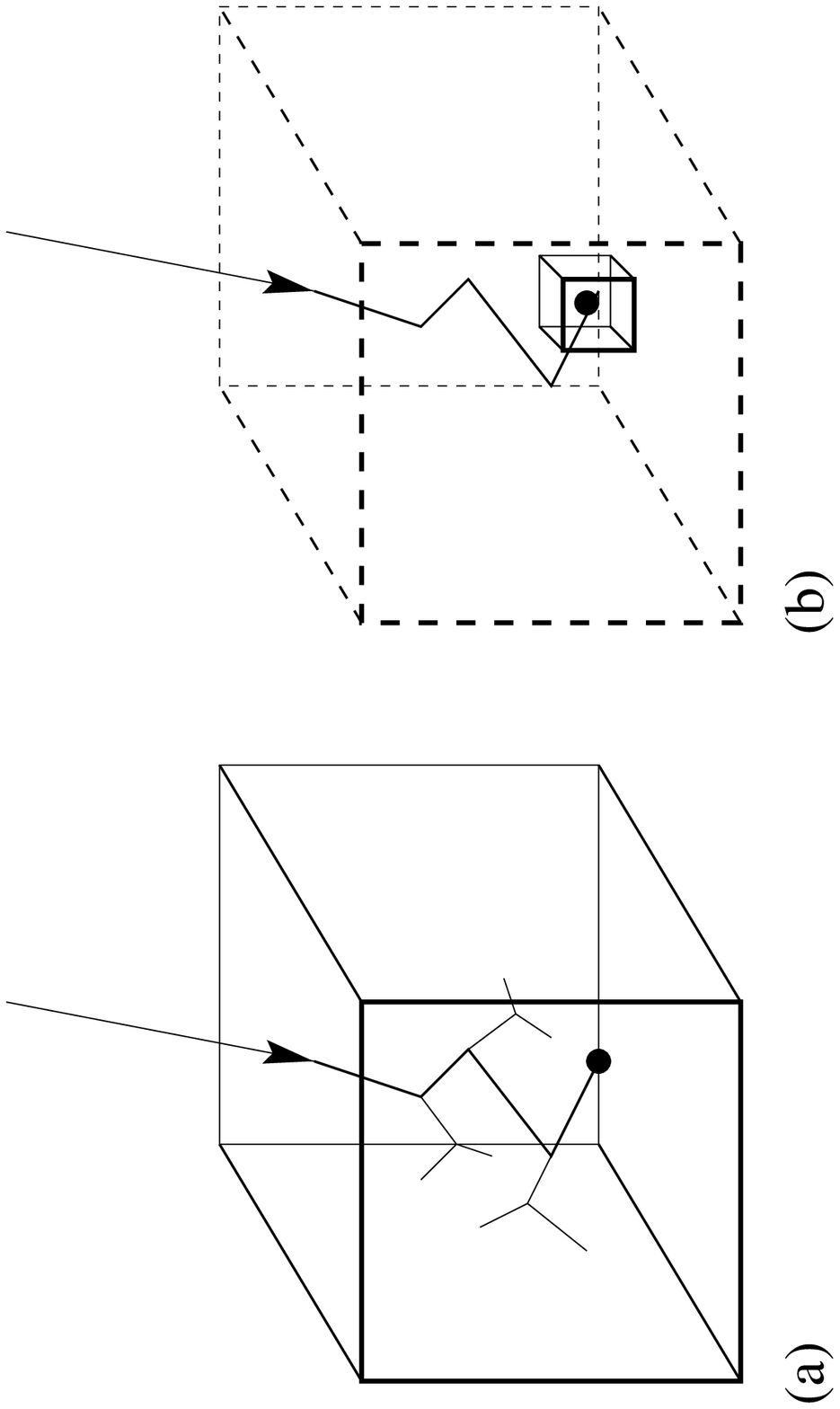}
\vspace{3.30 in}}

\caption{Schematic showing the relationship between (a) full MD,
and (b) the domain following approximation.}
\label{md_scheme}
\end{figure}

As we are only concerned with the path of the implanted ion,
we only need to consider the region of silicon immediately
surrounding the ion.
We continually create and destroy silicon atoms, to follow the domain of
the substrate that contains the ion. Material is built in slabs one
unit cell thick to ensure that the ion is always surrounded by a given
number of cells on each side. Material is destroyed if it is outside
the domain defined by the ion position and the domain thickness.
In this scenario, the ion sees the equivalent of a complete crystal,
but primary knock-on atoms (PKAs) and material in the wake of the ion
path behave unphysically, due to the small system dimensions.
Hence we have reduced the cost of the algorithm to O($u$),
at the expense of losing information on the final state of the Si substrate.
This algorithm is similar to the `translation' approach used in the
MDRANGE computer code developed by Nordlund\cite{nor95}.
The relationship between the full and restricted MD approaches
is shown in Fig.\ \ref{md_scheme}.

\begin{figure}
\vbox{\vspace{1.6 in} \hspace{0.4 in}
\includegraphics{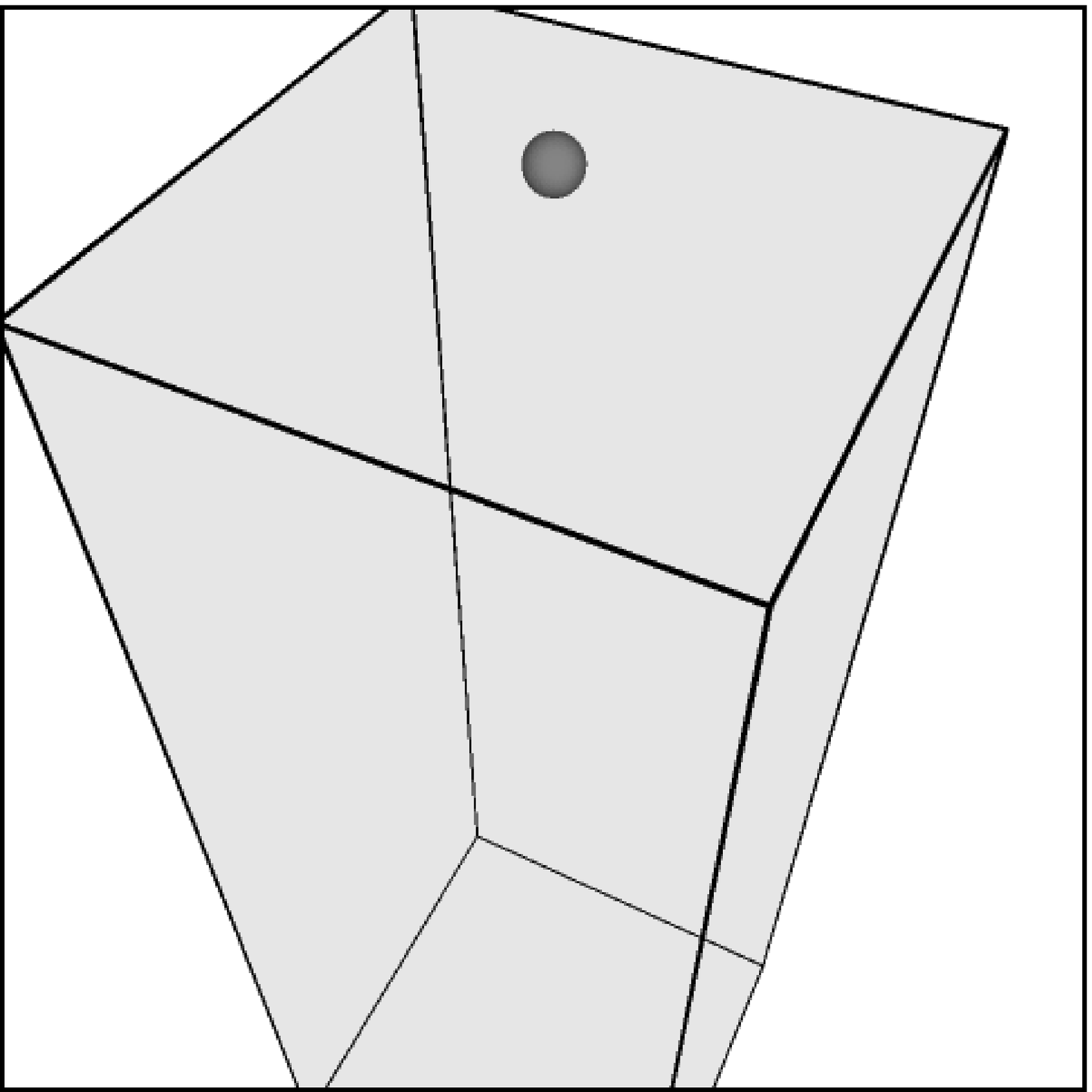}
\hspace{2.0 in}
\includegraphics{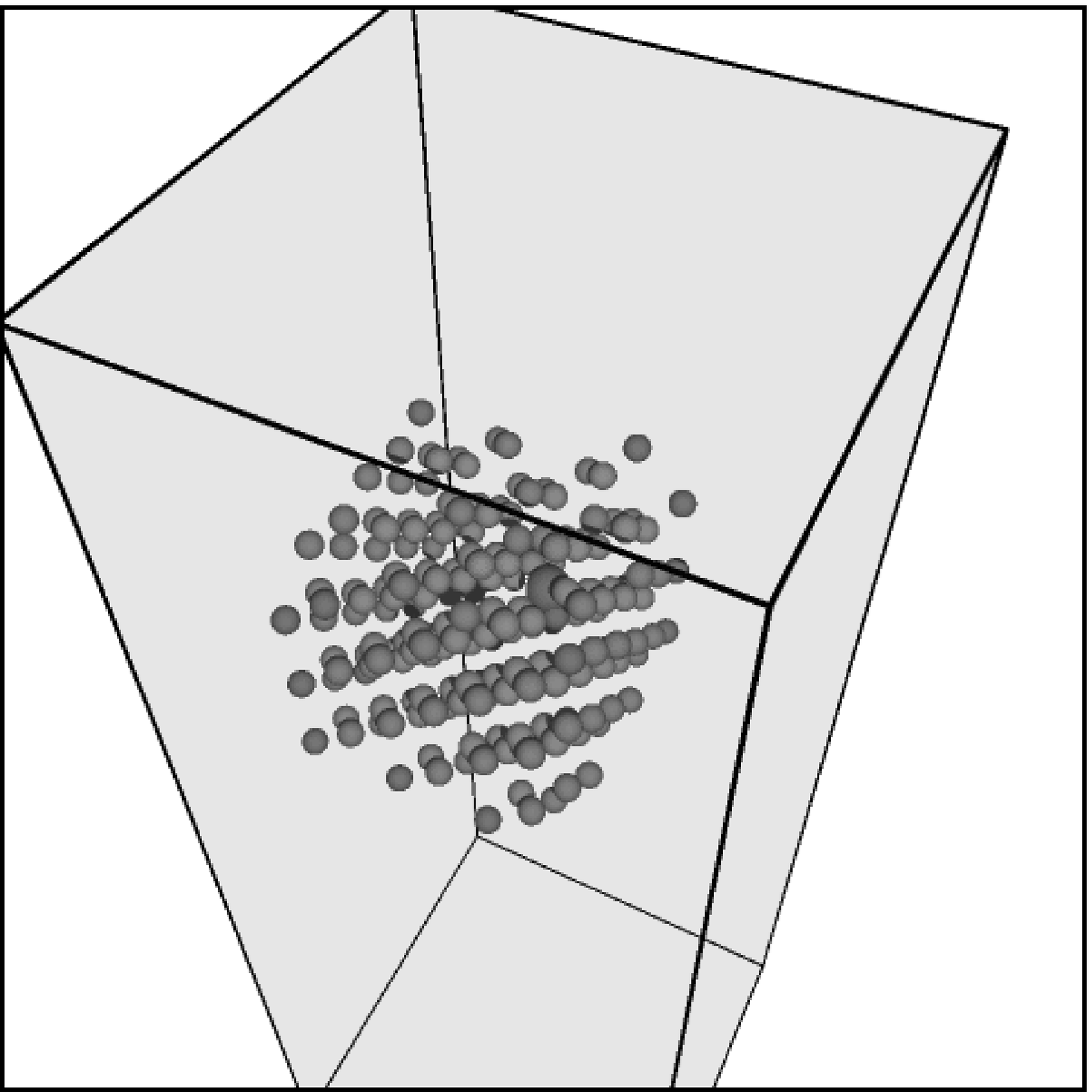}
\hspace{2.0 in}
\includegraphics{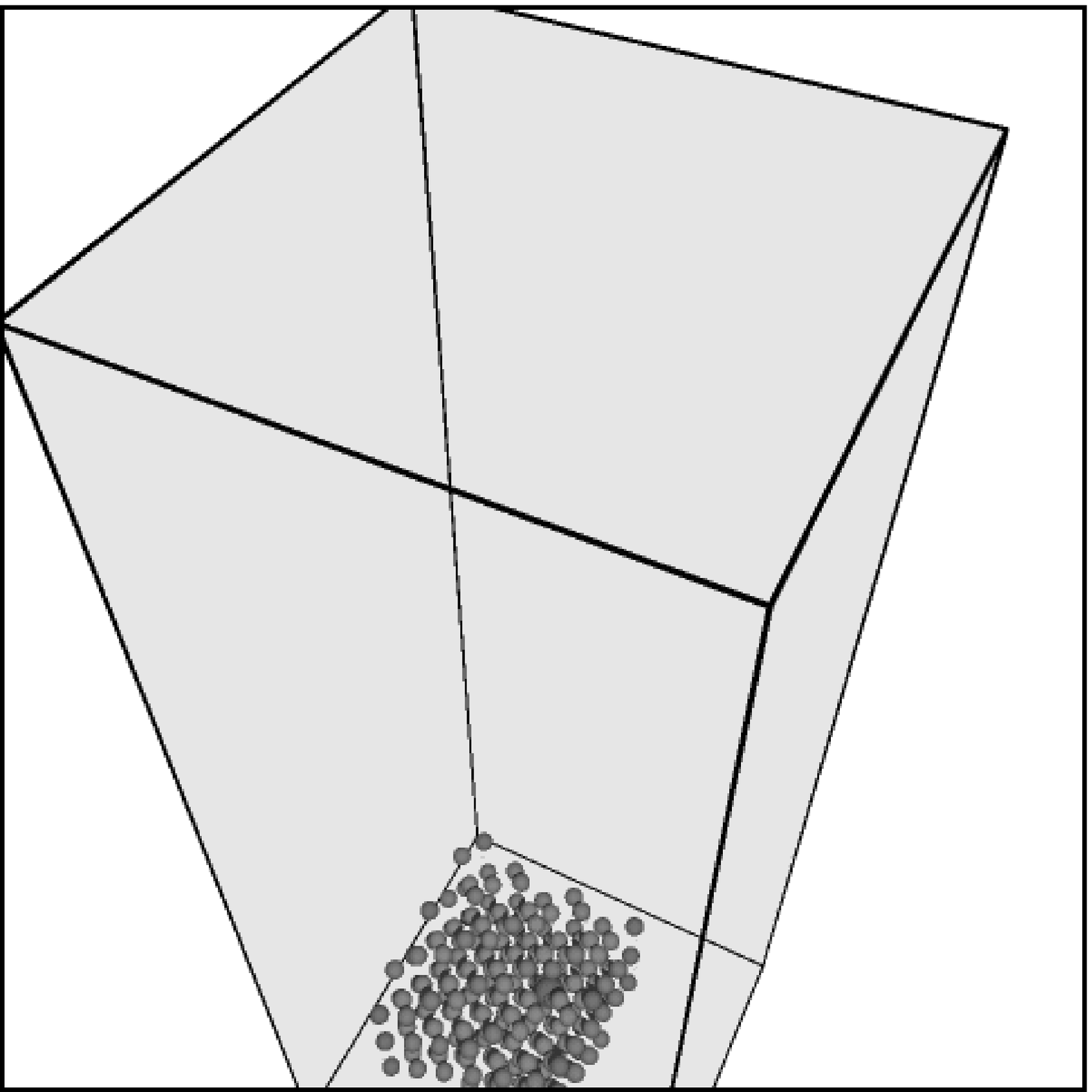}
\vspace{0.1 in}}

\caption{Initial stages during a domain following simulation;
the shaded box shows the crystal surface and orientation.}
\label{md_gifs}
\end{figure}

Fig.\ \ref{md_gifs} illustrates a single domain following trajectory.
The ion is initially above a semi-infinite volume that is the silicon target.
As the ion approaches the surface, atoms begin to be created in front of it,
and destroyed in its wake.
This process is continued until the ion comes to rest at some depth in the
silicon substrate. Several thousand of such trajectories are combined
to produce the depth profile of implanted ions.

\subsection{Moving Atom Approximation}
   This was first introduced by Harrison\cite{har88,smi89} to increase
the efficiency of ion sputtering yield simulations.
In this scheme atoms are divided
into two sets; those that are `on' have their positions integrated, and
those that are `off' are stationary. At the start of the simulation, only the
ion is turned on, and is the only atom to have forces calculated and to be
integrated. Some of the `off' atoms will be used in the force calculations
and will have forces assigned to them. If the resultant force
exceeds a certain threshold, the atom is turned on and its motion is
integrated. The simulation proceeds in this way with more and more atoms
having their position integrated as energy becomes dispersed
throughout the system.

   We use two thresholds in our simulation; one for atoms interacting
directly with the ion, and one for atom-atom interactions. We are,
of course, mostly concerned with generating the correct motion for the
ion, so the ion-atom interactions are of the most critical and
require a lower threshold than the atom-atom interactions.
In fact, for any reasonable threshold value,
almost any ion-atom interaction will result in the atom being turned on,
due to the large ion energy. Hence the ion-atom threshold is set to zero in
these simulations, as adjusting the value gives no increase in efficiency.

   In the case of the atom-atom threshold, we estimate a reasonable value
by comparison to simulations without the moving atom approximation (MAA).
Smith et al.\cite{smi89} found a force threshold of $1.12 \times 10^{-9}$ N
for both atom-atom and ion-atom interactions
gave the correct sputtering yield (when compared to simulations without
the MAA) in the case of 1 keV Ar implant into Si.
We have found a larger value ($8.0 \times 10^{-9}$ N) gives the correct
dopant profile, when compared to a simulations without the approximation.
Our ability to use a larger value is due to two reasons.
The motion of atoms not directly interacting with the ion only has a
secondary effect on its motion by influencing the position of directly
interacting atoms, so small errors in the positions of these atoms has
little consequence. Also, by dividing the interactions into two sets,
we do not have to lower the threshold to give the correct ion-atom
interactions.

\subsection{Pair Potential Approximation and Recoil Interaction Approximation}
While we use a many-body potential to describe a stable silicon
lattice for low energy implants, this introduces a significant
overhead to our simulations.
For higher ion velocities, we do not need to use such a level of detail.
A pair potential is sufficient to model the Si-Si interactions, as only
the repulsive interaction is significant. Also, as the
lattice is built at a metastable point with respect to a pair potential,
with atoms initially frozen due to the MAA, 
and the section of material is only simulated for a short period of time,
stability is not important. 
Hence, at a certain ion velocity we switch from the complete many-body
potential to a pair potential approximation (PPA) for the Si-Si
interactions. This is achieved in our code by setting the many-body
parameter within the Tersoff potential to its value for
undistorted tetrahedral Si, and results
in a Morse potential splined to a screened coulomb potential.

We make a further approximation for still higher ion energies, where only
the ion-Si interactions are significant in determining the ion path.
For ion velocities above a set
threshold we calculate only ion-Si interactions. This approximation,
termed the recoil interaction approximation (RIA)\cite{nor95},
brings the MD scheme close to many BCA implementations.
The major difference that exists between
the two approaches is that the ion path is obtained by integration, rather
than by the calculation of asymptotes, and that multiple interactions are,
by the nature of the method, handled in the correct manner.

We have determined thresholds of 90.0 eV/$m_{\text{u}}$ and 270.0 eV/$m_{\text{u}}$ for the PPA,
and RIA, respectively are sufficiently high that both
low and high energy calculated
profiles are unaffected by their use. As the thresholds are based on the
ion velocity, a single high energy ion simulation will switch between
levels of approximation as the ion slows down and will produce the correct
end of range behavior.

\section{Rare Event Algorithm}
\begin{figure}
\vbox{\vspace{-0.90 in} \hspace{0.7 in}
\includegraphics{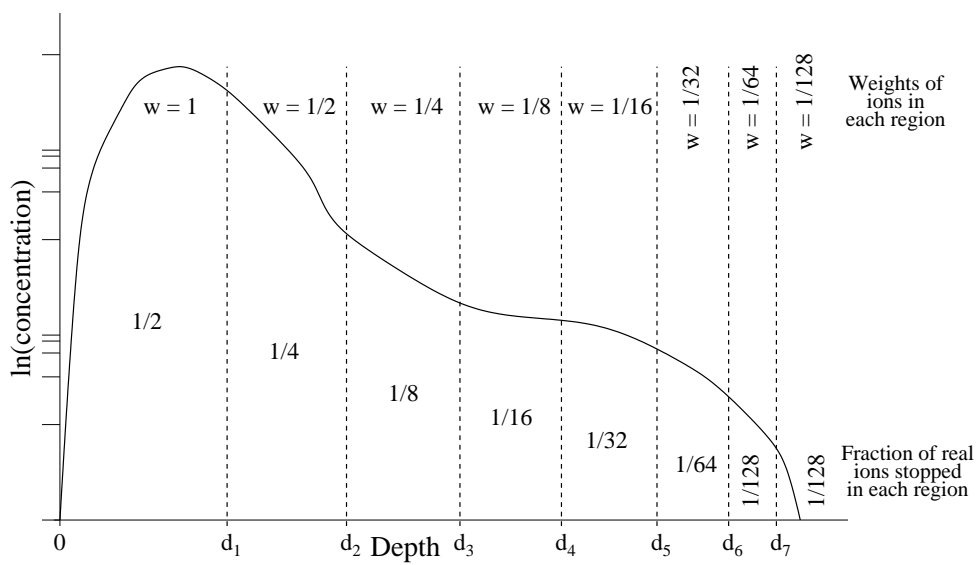}
\vspace{3.70 in}}

\caption{The algorithm for generating splitting depths from
the integrals of an existing dopant profile,
plus the weights associated with split ions at each depth.}
\label{splits}
\end{figure}

   A typical dopant concentration profile in crystalline silicon,
as illustrated in Fig.\ \ref{splits}, has a
characteristic shape consisting of a near-surface peak followed by an
almost exponential decay over some distance into the material, with a
distinct end of range distance.
The concentration of dopant in the tail of the profile is several orders of
magnitude less than that at the peak.
Hence if we wish to calculate a statistically significant concentration at
all depths of the profile we will have to run many ions that are stopped
near the peak for every one ion that stops in the tail, and most of
the computational effort will not enhance the accuracy of the profile
we are generating.

   In order to remove this redundancy from our calculations,
we employ an `atom splitting' scheme\cite{hub96}
to increase the sampling in the deep
component of the concentration profile. Every actual ion implanted is replaced
by several virtual ions, each with an associated weighting.
At certain \emph{splitting depths} in the material, each ion
is replaced by two ions, each with a weighting of half that prior to splitting.
Each split ion trajectory is run separately, and the weighting of
the ion is recorded along with its final depth. As the split ions see
different environments (material is built in front of the ion, with random
thermal displacements), the trajectories rapidly diverge from one another.
Due to this scheme, we can maintain the same number of virtual
ions at any depth, but their weights decrease with depth.
Each ion could of course be split into more than two at each depth, with
the inverse change in the weightings, but for simplicity and to keep the ion
density as constant as possible we work with two.

To maximize the advantages of this scheme, we dynamically update the
splitting depths. The correct distribution of splitting
depths is obtained from an approximate profile for the dopant
concentration.
 The initial profile is either read in (e.g. from SIMS
data), or estimated from the ion type, energy and incident direction using
a crude interpolation scheme based on known depths and concentrations
for the peak and tail. Once the simulation is running, the profile and
the splitting depths are re-evaluated at intervals.
The algorithm to determine the splitting depths from a given profile
is illustrated in Fig.\ \ref{splits}.
At the start of the simulation, we specify the number of orders of
magnitude, $M$, of change in the concentration of moving ions over which
we wish to reliably calculate the profile.
We split ions at depths where the total number of ions (ignoring weighting)
becomes half of the number of actual implanted ions. Hence we will
use $N$ splitting depths, where $N$ is the largest integer
$\le M \times \log_2 10$.
The splitting depths, $d_i$ ($1 \le i \le N$), are then chosen such that:
\begin{equation}
\int_{0}^{d_i} C(x)\, dx = ( 1 - {(\case{1}{2})}^i ) \times \int_{0}^{\infty} C(x)\, dx
\label{split}
\end{equation}
where $C(x)$ is the concentration of stopped ions
(i.e., the dopant concentration) at depth $x$.
Although we are using an approximate profile from few ions to
generate the splitting
depths, the integration is a smoothing operation and so
gives good estimates of the splitting depths.

To minimize the storage requirements due to ion splitting, each real ion
is run until it comes to rest,
and the state of the domain is recorded at each splitting depth passed.
The deepest split ion is then run, and further split ions are stored if
it passes any splitting depths.
This is repeated until all split ions have been run, then the next real
ion is started. Hence the maximum we ever need to store is one domain per
splitting depth (i.e., 16 domains when splitting over 5 orders of magnitude).

\section{Simulation Details}
All simulations were run with a Si\{100\} target at a temperature of 300 K.
A surface amorphous layer of one, or three unit cells thickness was used.
Dopant profiles were calculated for As, B, P, and Al ions;
in each case it was assumed that only the most abundant
isotope was present in the ion beam.
The direction of the incident ion beam is specified by the angle
of tilt, $\theta^{\circ}$, from normal and the azimuthal angle $\phi^{\circ}$,
as ($\theta$,$\phi$). 
The incident direction of the ions was
either (0,0), i.e. normal to the surface
($\langle$100$\rangle$ channeling case),
(7--10,0--30) (non-channeling),
or (45,45) ($\langle$110$\rangle$ channeling),
and a beam divergence of 1.0$^{\circ}$ was always assumed.
    Simulations were run for 1,000 ions, with the splitting depths
updated every 100 ions.
A domain of 3$\times$3$\times$3 unit cells was used and
the profile was calculated over either 3, or 5 orders of
magnitude change in concentration.
The simulations were run on
pentium pro workstations running the Red Hat Linux operating system
with the GNU g77 Fortran compiler, or SUN Ultra-sparc workstations
with the SUN Solaris operating system and SUN Fortran compiler.
The running code typically requires about 750K of memory.

\section{Results and Discussion}
Two sets of results are presented; we first demonstrate the effectiveness
and stability of the rare event enhancement scheme, and then give examples
of data produced by the simulations and compare to SIMS data. Example
timings from simulations are also given.
A more extensive set of calculated profiles will be published
separately \cite{bea97b}.

\subsection{Performance of the Rare Event Algorithm}
\begin{figure}
\vbox{\vspace{-0.70 in} \hspace{0.7 in}
\includegraphics{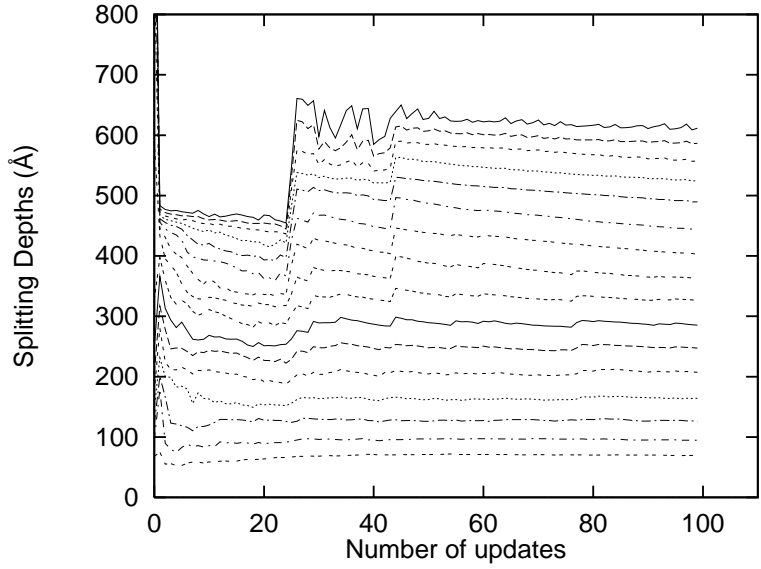}
\vspace{3.50 in}}

\caption{The evolution of splitting depths during the course of a simulation
(1,000 5 keV As ions (10,22) into Si\{100\}, updates every 10 real ions).}
\label{As10t22r5Ks}
\end{figure}

An example of the evolution of splitting depths during a simulation
is shown in Fig.\ \ref{As10t22r5Ks}, for the case of non-channeling
5 keV As implanted into Si\{100\}. The positions of the splitting depths
near the peak stabilize quickly. Splitting depths near the tail take far
longer to stabilize, as these depend on ions that channel the maximum
distance into the material. Although atom splitting enhances the number of
(virtual) ions that penetrate deep into the material, the occurrence of an
ion that will split to yield ions at these depths is still
a relatively rare event. The fact that all splitting depths do stabilize
is also an indication that we have run enough ions to generate good
statistics for the entire profile.

\begin{figure}
\vbox{\vspace{-3.1 in} \hspace{0.9 in}
\includegraphics{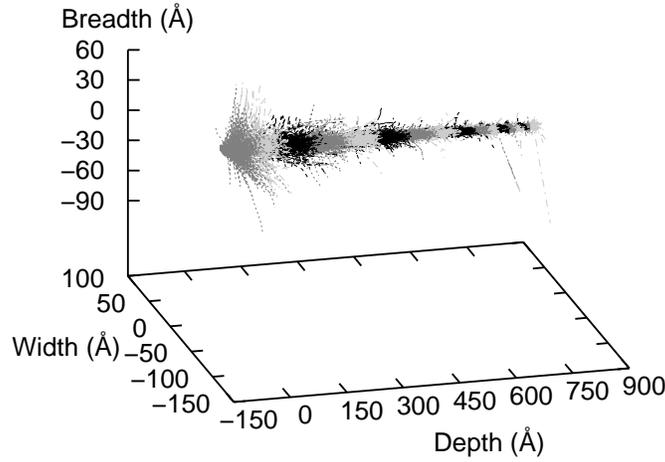}
\vspace{5.20 in}}

\caption{
The paths of 1,000 5 keV As ions implanted into Si\{100\}
at normal incidence, simulated in order to calculate the dopant
concentration profile over 5 orders of magnitude.
The paths of all split ions are shown,
shaded to show the number of times they were split.}
\label{psplit}

\end{figure}

\begin{figure}
\vbox{\vspace{-3.2 in} \hspace{0.9 in}
\includegraphics{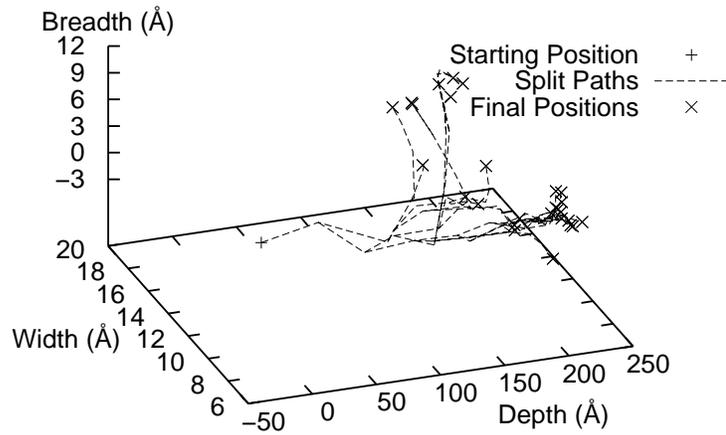}
\vspace{5.20 in}}

\caption{
The paths of virtual ions due to the implant of one real
5 keV As ion into Si\{100\} at normal incidence.}
\label{osplit}

\end{figure}

\begin{figure}
\vbox{\vspace{-2.7 in} \hspace{0.5 in}
\includegraphics{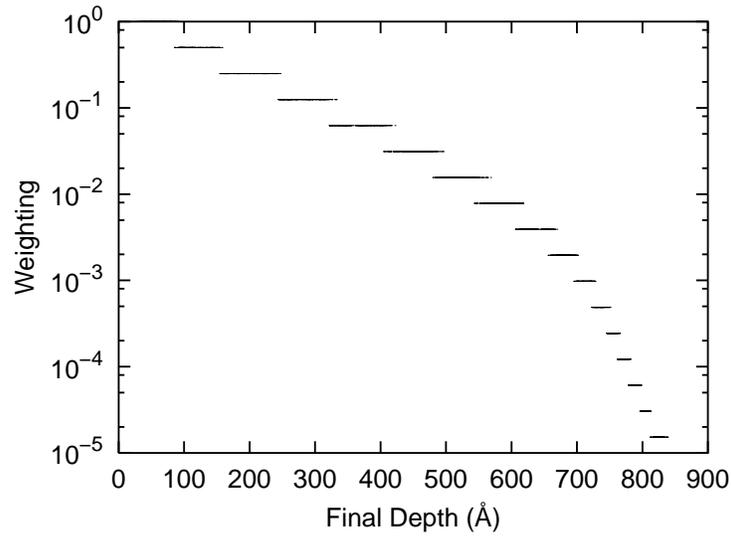}
\vspace{5.30 in}}

\caption{
The weighting of the last 500 of 1,000 5 keV As ions implanted into Si\{100\}
at normal incidence, plotted against final depth.}
\label{wsplit}

\end{figure}

The paths of 5 keV As ions implanted at normal incidence into Si\{100\}
are shown in Fig.\ \ref{psplit},
with the number of splittings shown by the line shading.
The 1,000 implanted real ions were split to yield a total
of 19,270 virtual ions.
The paths taken by 27 split ions produced from the first real ion of
this simulation,
and the resulting distribution of the ion positions
are shown in Fig.\ \ref{osplit}.
The final ions are the result of between 3 and 6 splittings,
depending upon the range of each trajectory. This is typical of
the distribution of splittings for one real ion; the final depths of ions
are not evenly distributed over the entire ion range, but are bunched
around some point within this range. This reflects how the impact position
of the ion and collisions during its passage through the amorphous layer
affect its ability to drop into a channel once in the crystalline material.
The weighting of the second 500 of the ions
(after the splitting depths had stabilized) is plotted against final depth
in Fig.\ \ref{wsplit} (note the log scale).

\begin{figure}
\vbox{\vspace{-2.7 in} \hspace{-1.5 in}
\includegraphics{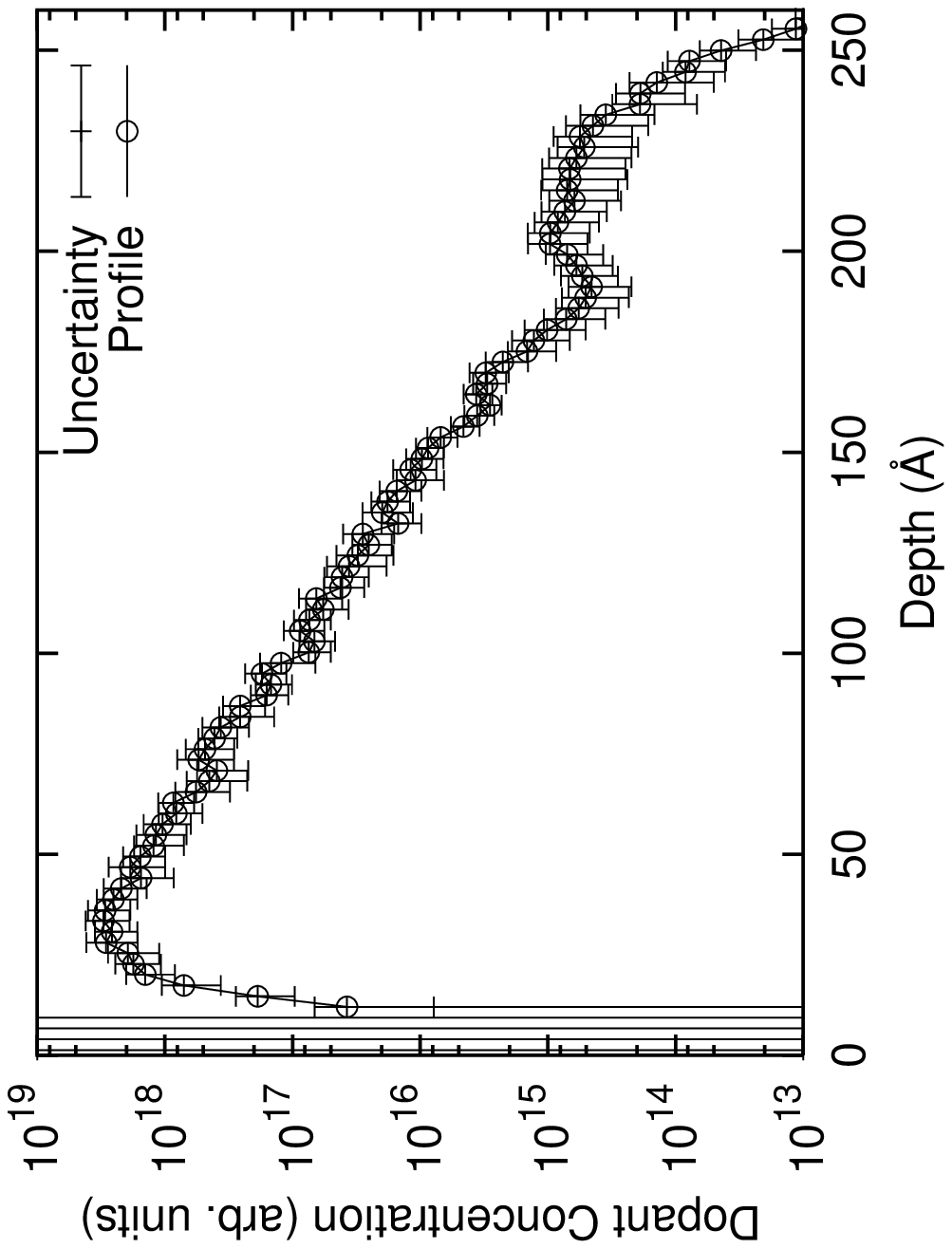}
\vspace{-0.15 in}
\hspace{3.6 in}
\includegraphics{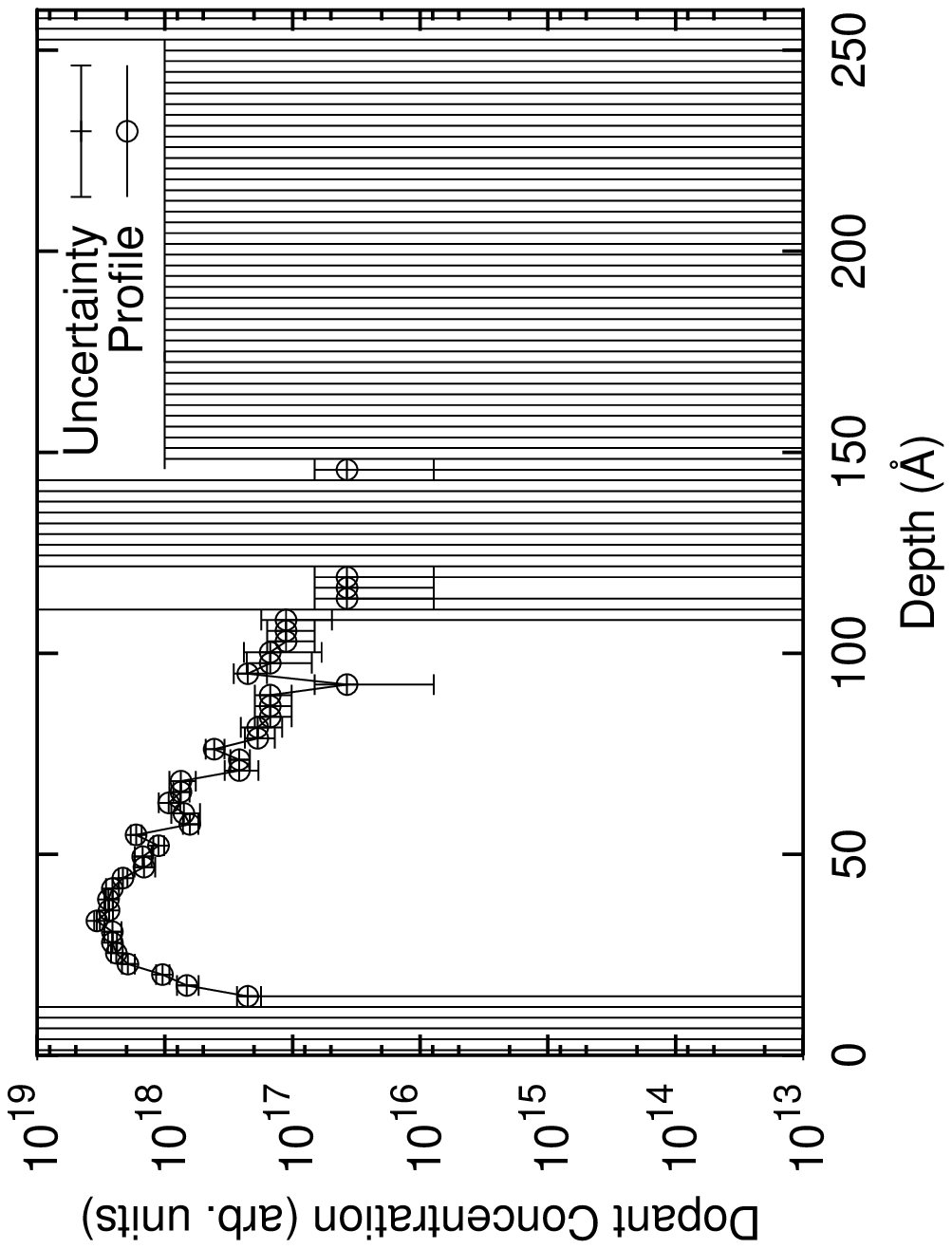}
\vspace{5.40 in}}

\caption{
The estimated uncertainty in the calculated dopant profile due to
2 keV As (7,0) into Si\{100\},
for the same number of initial ions (1,000),
with and without rare event enhancement.}
\label{var}

\end{figure}

We have estimated the uncertainty in the calculated dopant profiles
in order to judge the increase in efficiency obtained through the use
of the rare event enhancement scheme.
The uncertainty was estimated by dividing the final ion depths into 10 sets.
A depth profile was calculated from each set using a histogram of 100 bins,
with constant bin size.
A reasonable measure of the uncertainty is the standard deviation of the
distribution of the 10 concentrations for each bin.
   Fig.\ \ref{var} shows calculated dopant profiles from 1,000 real ions
for the case of
2 keV As at (7,0) into Si\{100\}, obtained with
and without atom splitting over five orders of magnitude.
The profiles are plotted
with the uncertainty represented by the size of error bars;
the length of each error bar corresponds to half the standard deviation
of concentrations in that bin.
The uncertainty is constant in the case of the 
profile obtained with the rare event scheme,
whereas the profile obtained without the scheme is only reliable over
one order of magnitude.

   Timings from these simulations, and a simulation with splitting to
three orders of magnitude are given in Table~\ref{rare}. From these
timings, we can estimate the efficiency gain due to the rare event algorithm.
We decrease the time required by a factor of 89 in the case
of calculating a profile to three orders of magnitude, and by a factor of
886 when calculating a profile over 5 orders of magnitude, compared to the
estimated time requirements without rare event enhancement. The gain in
efficiency increases exponentially with the number of orders of magnitude
in concentration over which we wish to calculate the profile.

\subsection{Comparison of Profiles to SIMS Data}
        The remaining figures show the calculated concentration profile of
B, As, and P ions for various incident energies and directions.
Profiles were generated from a histogram of 100 bins,
using adaptive bin sizes; the final ion depths were sorted and the same
number of virtual ions assigned to each bin. No other processing,
or smoothing of the profiles was done.
Also shown are low dose ($\le$ $10^{13}$ ions/cm$^2$) SIMS data;
for comparison, all profiles were scaled to an effective dose
of $10^{12}$ ions/cm$^2$.
We have also examined Al ion implants, but were
unable to match calculated profiles to the available
SIMS data\cite{wil86} for a physically
reasonable parameter value in our electronic stopping model.
This may be due to one or more of the following reasons;
Al is the only metal that we are implanting; the Al-Si interaction is
the only interaction for which we do not have a pair specific ZBL potential;
we only have a very limited set of SIMS data to compare to.

   In the case of the low energy ($\le$ 10 keV) implants, we compare to SIMS
data obtained with a thin and well controlled surface layer\cite{bou91,tas97};
here we assume one unit cell thickness of surface disorder
in our simulations. For the other cases considered here\cite{tas97,sch91},
the surface was
less well characterized; we assume three unit cells of disorder at
the surface, as this is typical of implanted Si.
For the low energy implants, we have calculated
profiles over a change of five orders of magnitude in concentration; for
the higher energy implants we calculate profiles over 3 orders of magnitude.
       The results of the REED calculations show good agreement with the
experimental data. In the case of the low energy implants, the
SIMS profile is only resolved over two orders of magnitude in some cases,
while we can calculate the profile over five orders of magnitude.

We give timing results from several simulations, as examples
of the cpu requirements of our implementation of the model.
Note, the results presented here are from a functional version of REED, but
the code has yet to be fully optimized to take advantage of the small
system sizes (around 200 atoms).
Timing data are given in Table~\ref{time}, for profiles calculated over
five orders of magnitude on a single pentium pro.
Run times are dependent on
the ion type and its incident direction, but are most strongly linked to
the ion velocity. We estimate a runtime of approximately 30 hours per
$\sqrt{{\text{keV}}/m_{\text{u}}}$, for this version of our code.

\section{Conclusions}
  In summary,
we have developed a restricted molecular dynamics code to simulate the
ion implant process and directly calculate `as implanted' dopant profiles.
This gives us the
accuracy obtained by time integrating atom paths, whilst obtaining an
efficiency far in excess of full MD simulation.
        There is very good agreement between the MD results and SIMS data
for B, P, and As implants. We are unable to reproduce published SIMS data for
Al implants\cite{wil86} with our current model.
This discrepancy is currently
being investigated; our findings will be published separately\cite{cai97a}.
We can calculate the dopant profile to concentrations one or two orders of
magnitude below that measurable by SIMS for the channeling tail of low
dose implants.

The scheme described here gives a viable alternative to the BCA approach.
Although it is still more expensive computationally, it is sufficiently
efficient to be used on modern desktop computer workstations.
The method has two major advantages over the BCA approach:
(i) Our MD model consists only of standard empirical
potentials developed for bulk Si and for ion-solid interactions.
The only fitting is in the electronic stopping model, and
this involves \emph{only one} parameter per ion species.
This should be contrasted to the many parameters that
have to be fit in BCA models.
   We believe that by using physically based models for all aspects
of the program, with the minimum of fitting parameters, we obtain good
transferability to the modeling of implants outside of our fitting set.
(ii) The method does not break down at the low ion energies necessary for
production of the next generation of computer technology; it gives the
correct description of multiple, soft interactions that occur both
in low energy implants, and high energy channeling.

We are currently working to fully optimize the code, in order to
maximize its efficiency. The program is also being extended to include
a model for ion induced damage, amorphous and polycrystalline targets, and
to model cluster implants such as BF$_2$. We also note that the scheme can
be easily extended to include other ion species such as Ge, In and Sb,
and substrates such as GaAs and SiC.

\section{Acknowledgments}
  We gratefully acknowledge David Cai and Charles Snell
for providing us with their insight during many discussions,
and Al Tasch and co-workers for providing preprints of their work and
SIMS data.
This work was performed under the auspices of the United States Department
of Energy.

\section{References}

\protect\pagebreak
\section{Figures}
%
%

\begin{figure}
\vbox{\vspace{-2.7 in} \hspace{0.5 in}
\includegraphics{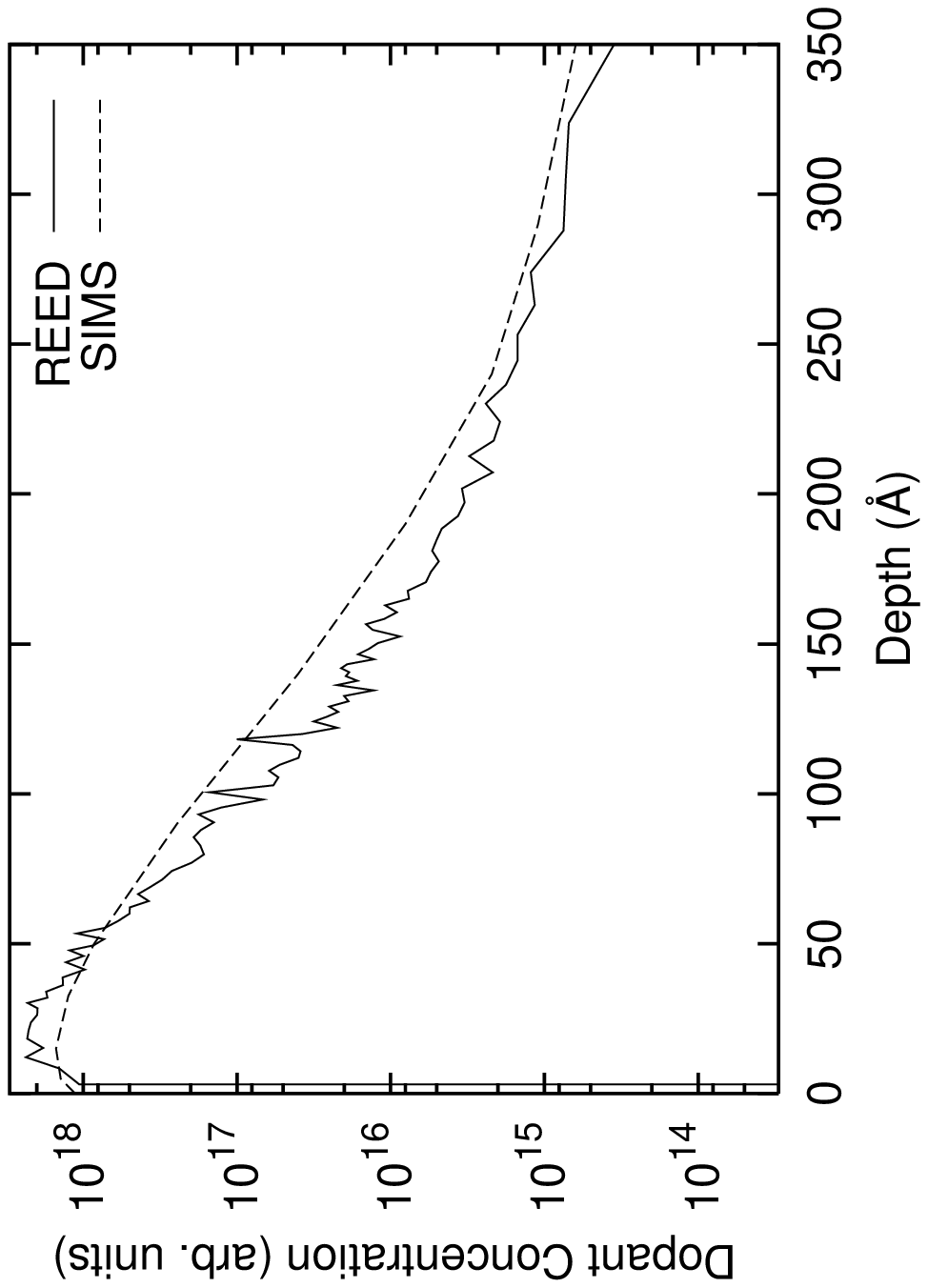}
\vspace{5.20 in}}

\caption{
The calculated and experimental\protect\cite{bou91} dopant profiles due to
0.5 keV B (0,0) into Si\{100\}.}
\label{B0t0r500}

\end{figure}

\begin{figure}
\vbox{\vspace{-2.7 in} \hspace{0.5 in}
\includegraphics{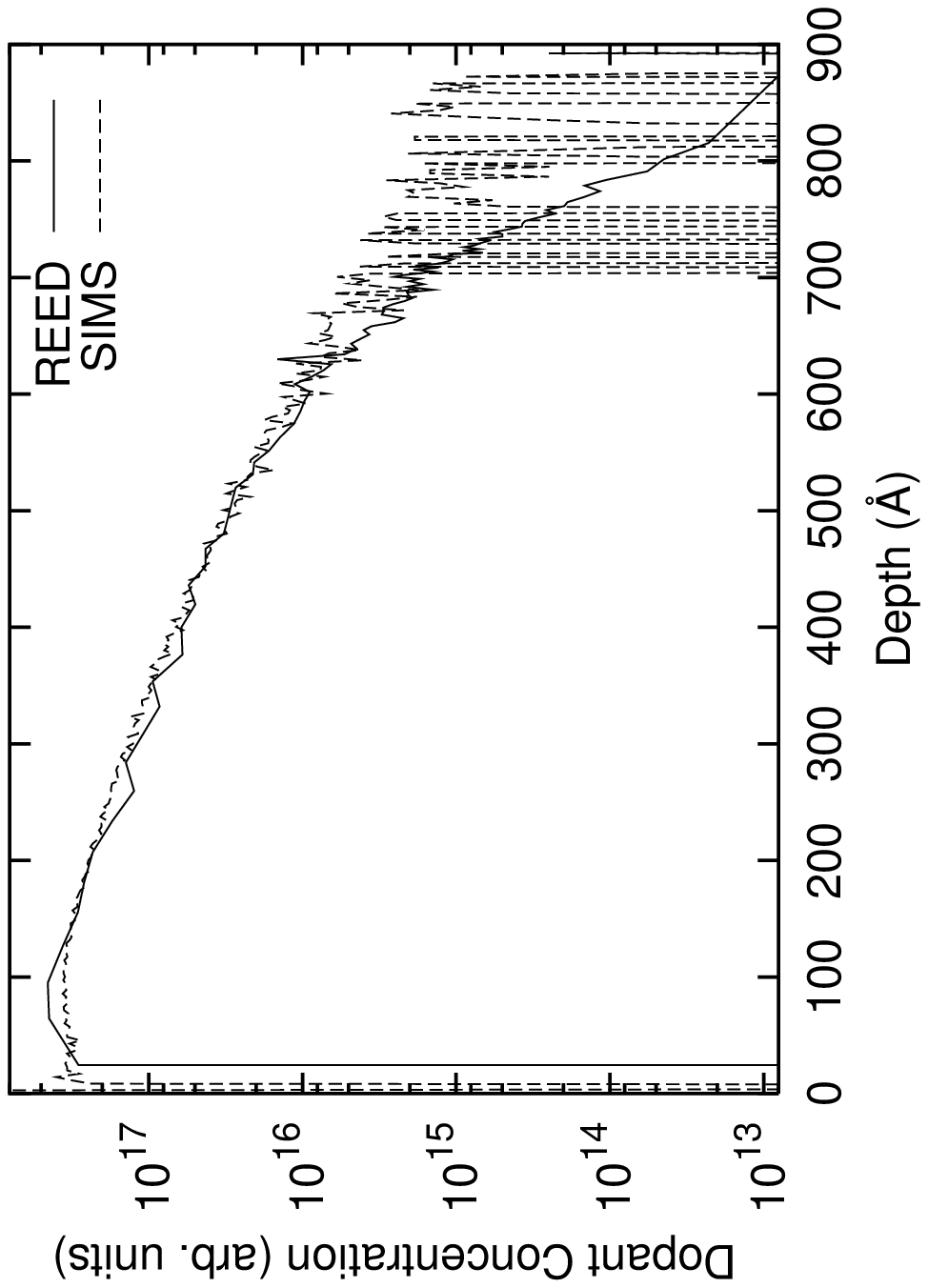}
\vspace{5.20 in}}

\caption{
The calculated and experimental\protect\cite{tas97} dopant profiles due to
2 keV B (0,0) into Si\{100\}.}
\label{B0t0r2K}

\end{figure}

\begin{figure}
\vbox{\vspace{-2.7 in} \hspace{0.5 in}
\includegraphics{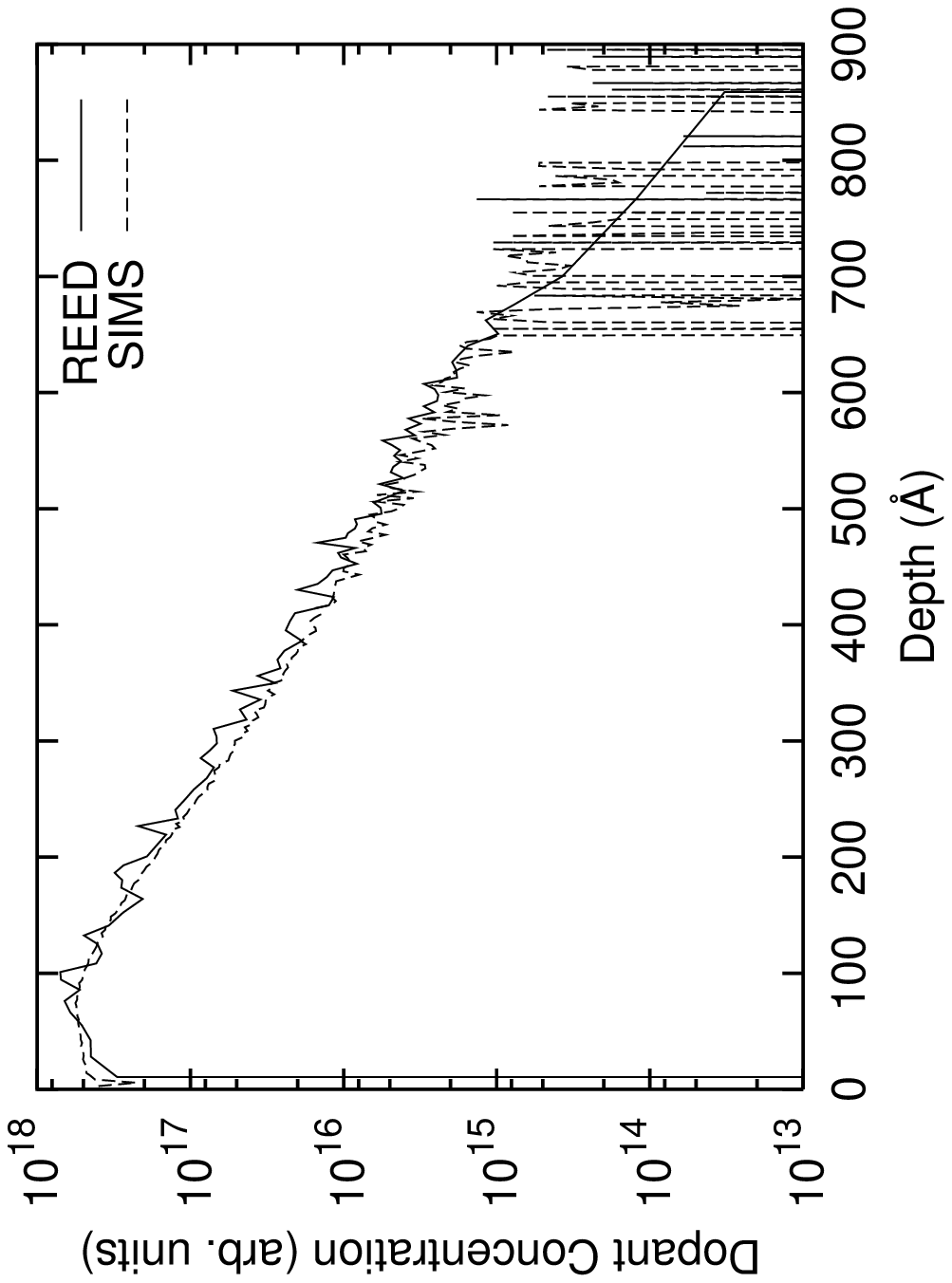}
\vspace{5.20 in}}

\caption{
The calculated and experimental\protect\cite{tas97} dopant profiles due to
2 keV B (7,0) into Si\{100\}.}
\label{B7t0r2K}

\end{figure}

\begin{figure}
\vbox{\vspace{-2.7 in} \hspace{0.5 in}
\includegraphics{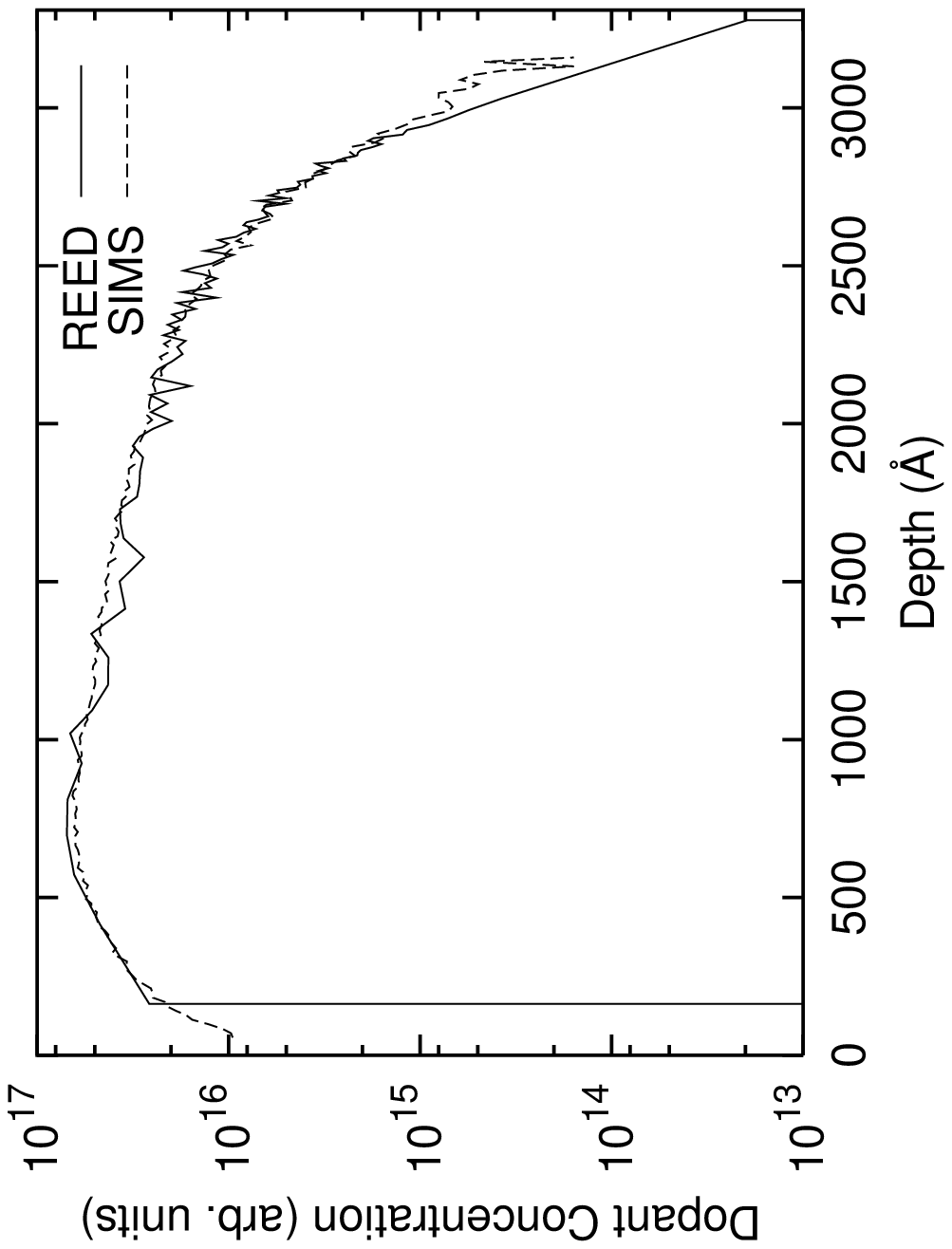}
\vspace{5.20 in}}

\caption{
The calculated and experimental\protect\cite{tas97} dopant profiles due to
15 keV B (0,0) into Si\{100\}.}
\label{B0t0r15K}

\end{figure}

\begin{figure}
\vbox{\vspace{-2.7 in} \hspace{0.5 in}
\includegraphics{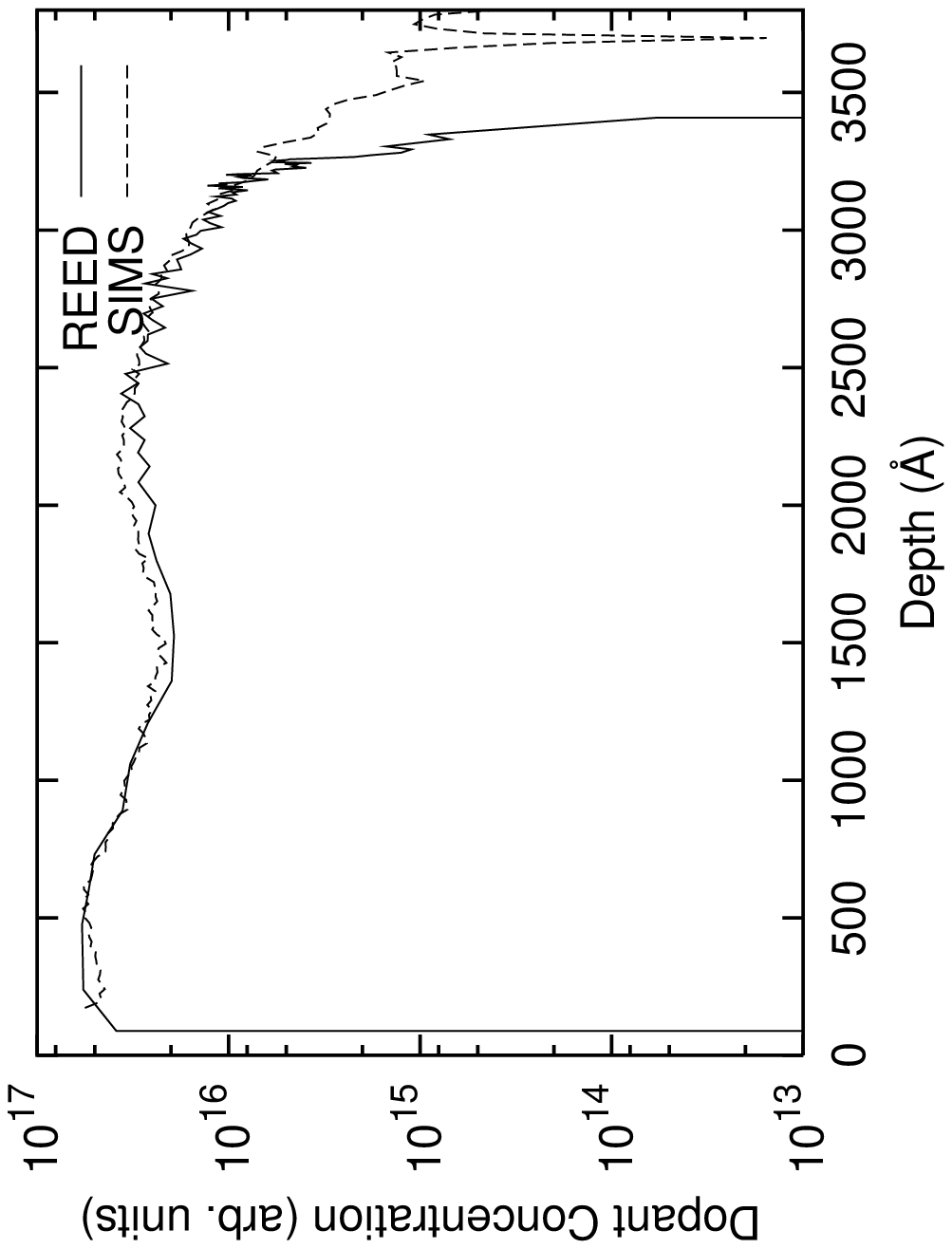}
\vspace{5.20 in}}

\caption{
The calculated and experimental\protect\cite{tas97} dopant profiles due to
15 keV B (45,45) into Si\{100\}.}
\label{B45t45r15K}

\end{figure}

\begin{figure}
\vbox{\vspace{-2.7 in} \hspace{0.5 in}
\includegraphics{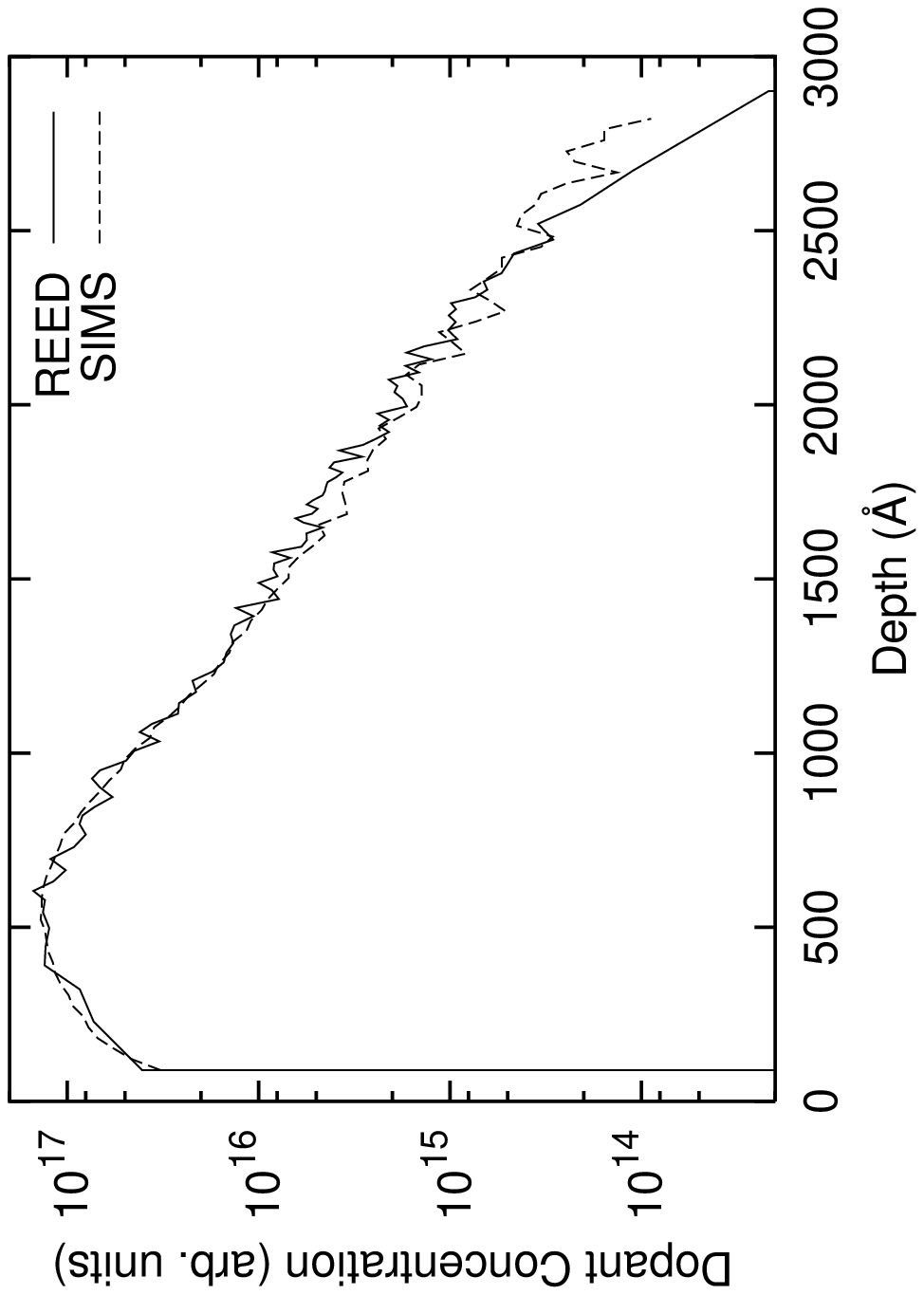}
\vspace{5.20 in}}

\caption{
The calculated and experimental\protect\cite{tas97} dopant profiles due to
15 keV B (7,30) into Si\{100\}.}
\label{B7t30r15K}

\end{figure}

\begin{figure}
\vbox{\vspace{-2.7 in} \hspace{0.5 in}
\includegraphics{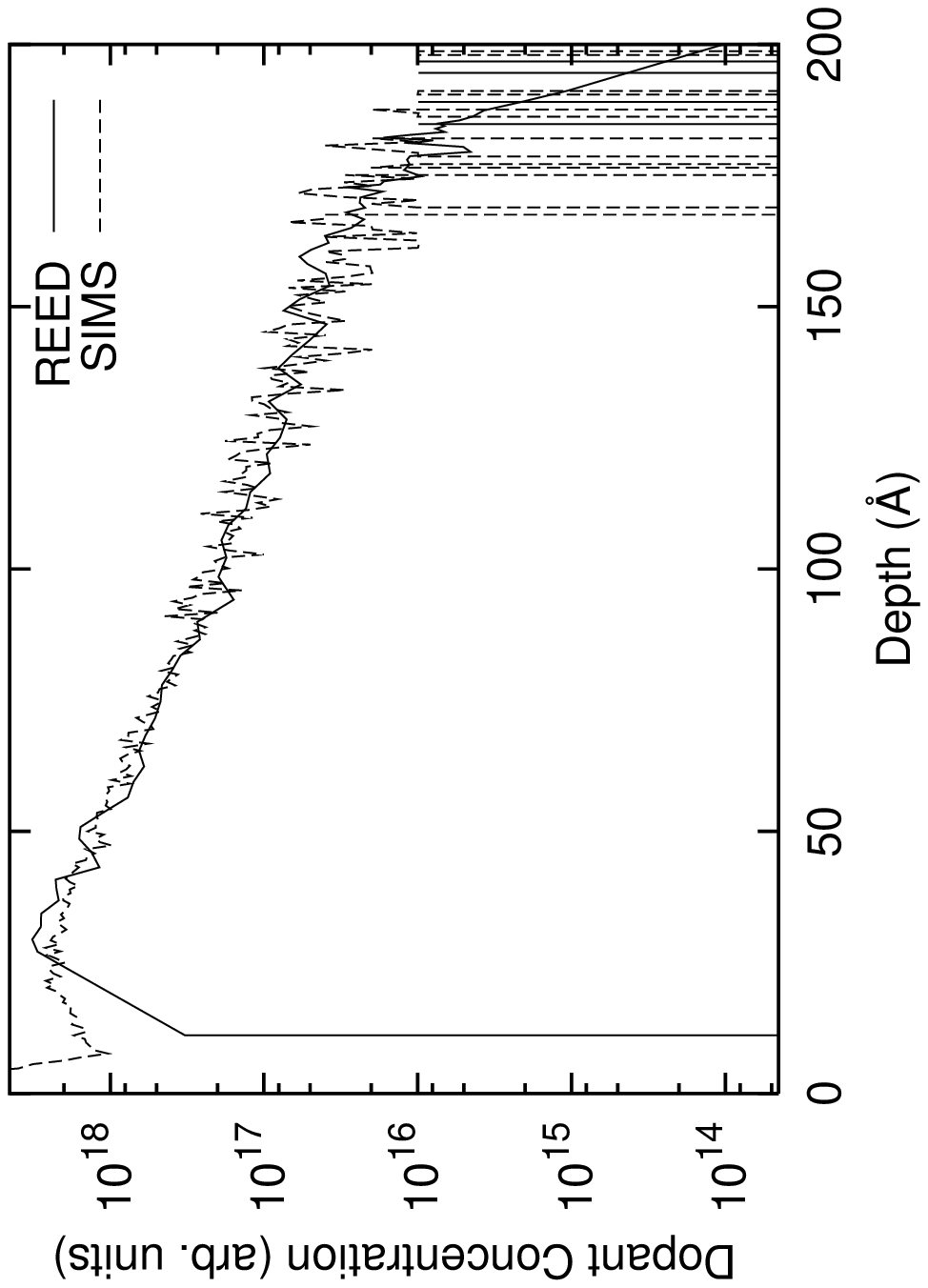}
\vspace{5.20 in}}

\caption{
The calculated and experimental\protect\cite{tas97} dopant profiles due to
2 keV As (0,0) into Si\{100\}.}
\label{As0t0r2K}

\end{figure}

\begin{figure}
\vbox{\vspace{-2.7 in} \hspace{0.5 in}
\includegraphics{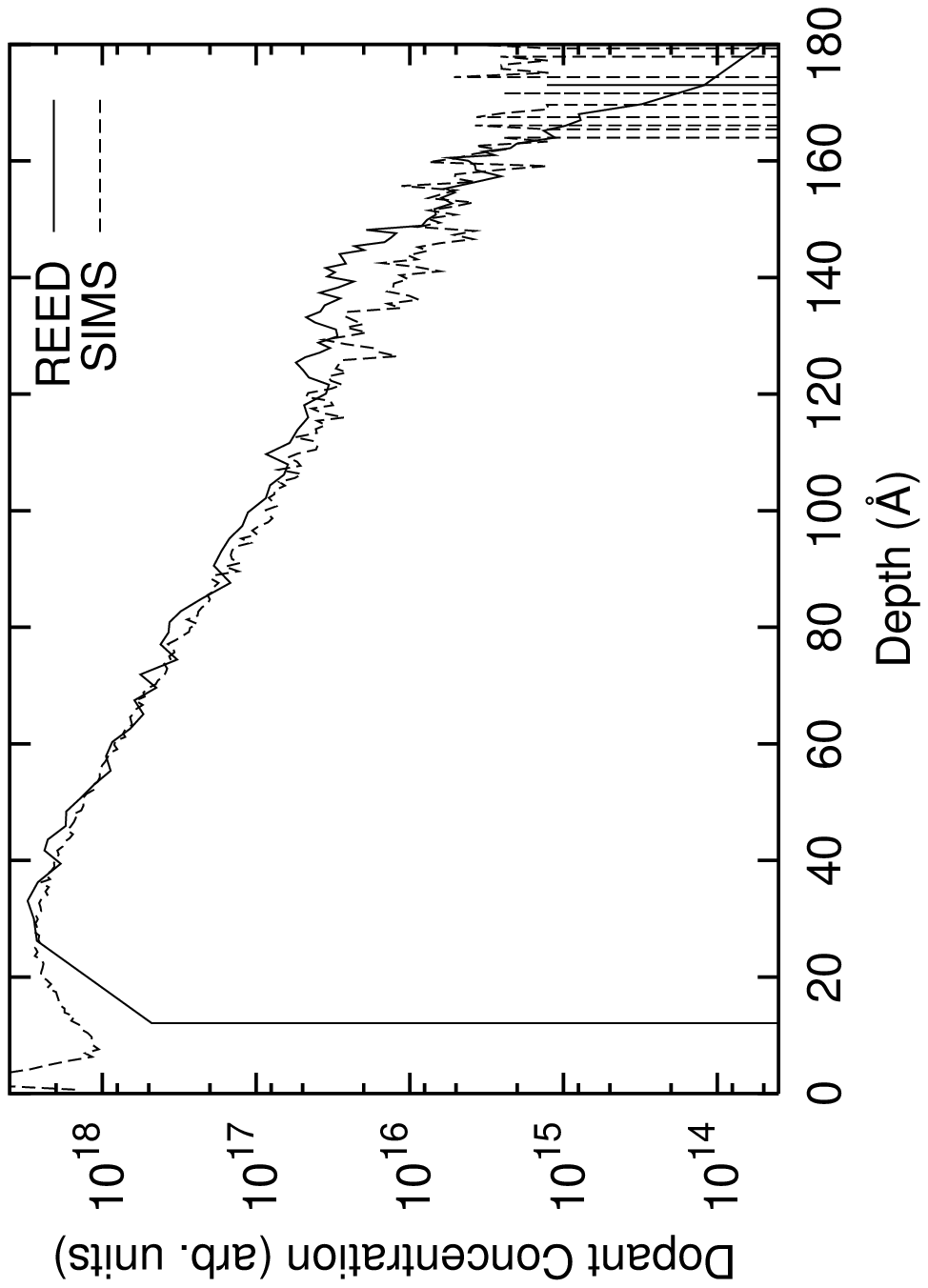}
\vspace{5.20 in}}

\caption{
The calculated and experimental\protect\cite{tas97} dopant profiles due to
2 keV As (7,0) into Si\{100\}.}
\label{As7t0r2K}

\end{figure}

\begin{figure}
\vbox{\vspace{-2.7 in} \hspace{0.5 in}
\includegraphics{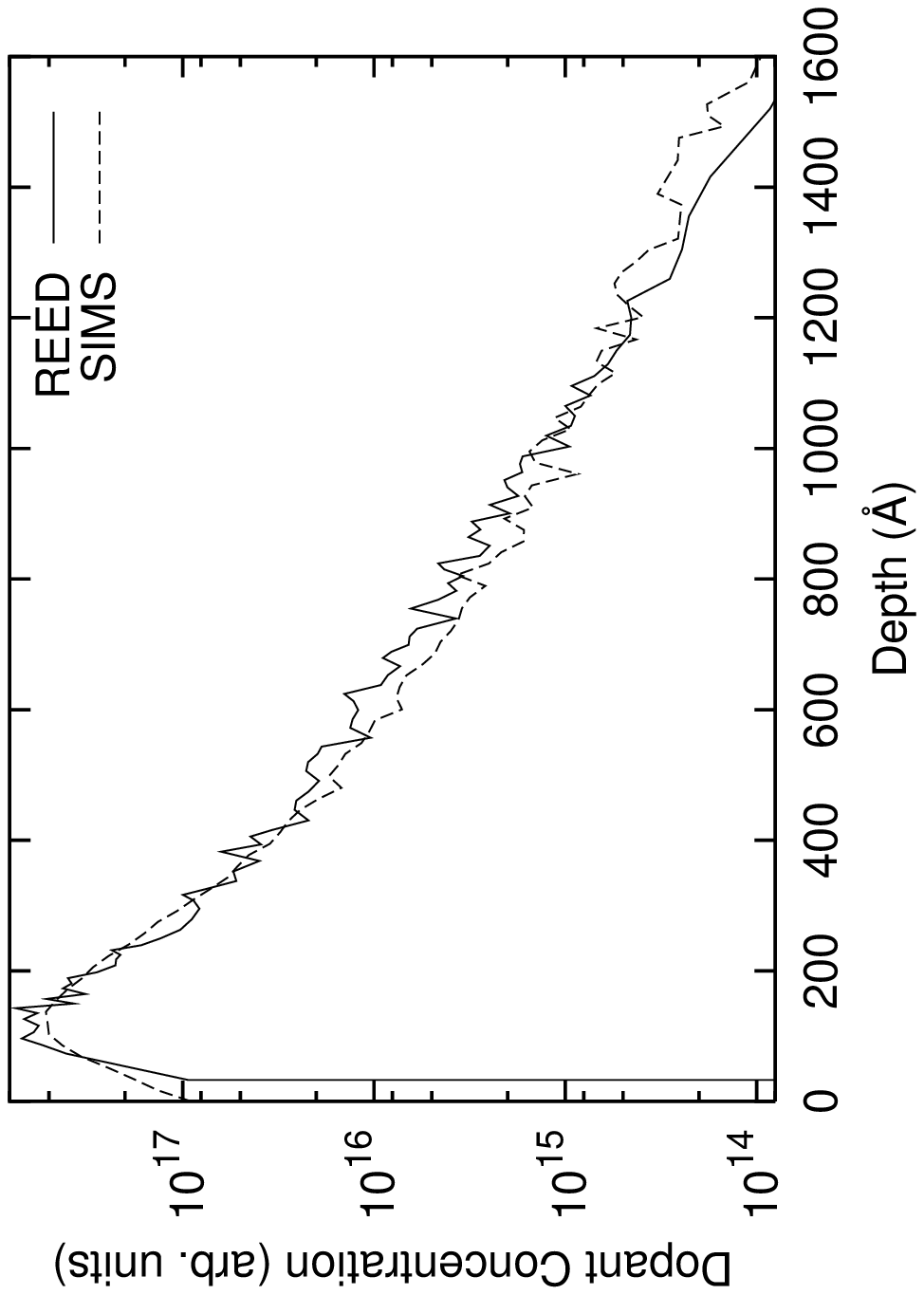}
\vspace{5.20 in}}

\caption{
The calculated and experimental\protect\cite{tas97} dopant profiles due to
15 keV As (8,30) into Si\{100\}.}
\label{As8t30r15K}

\end{figure}

\begin{figure}
\vbox{\vspace{-2.7 in} \hspace{0.5 in}
\includegraphics{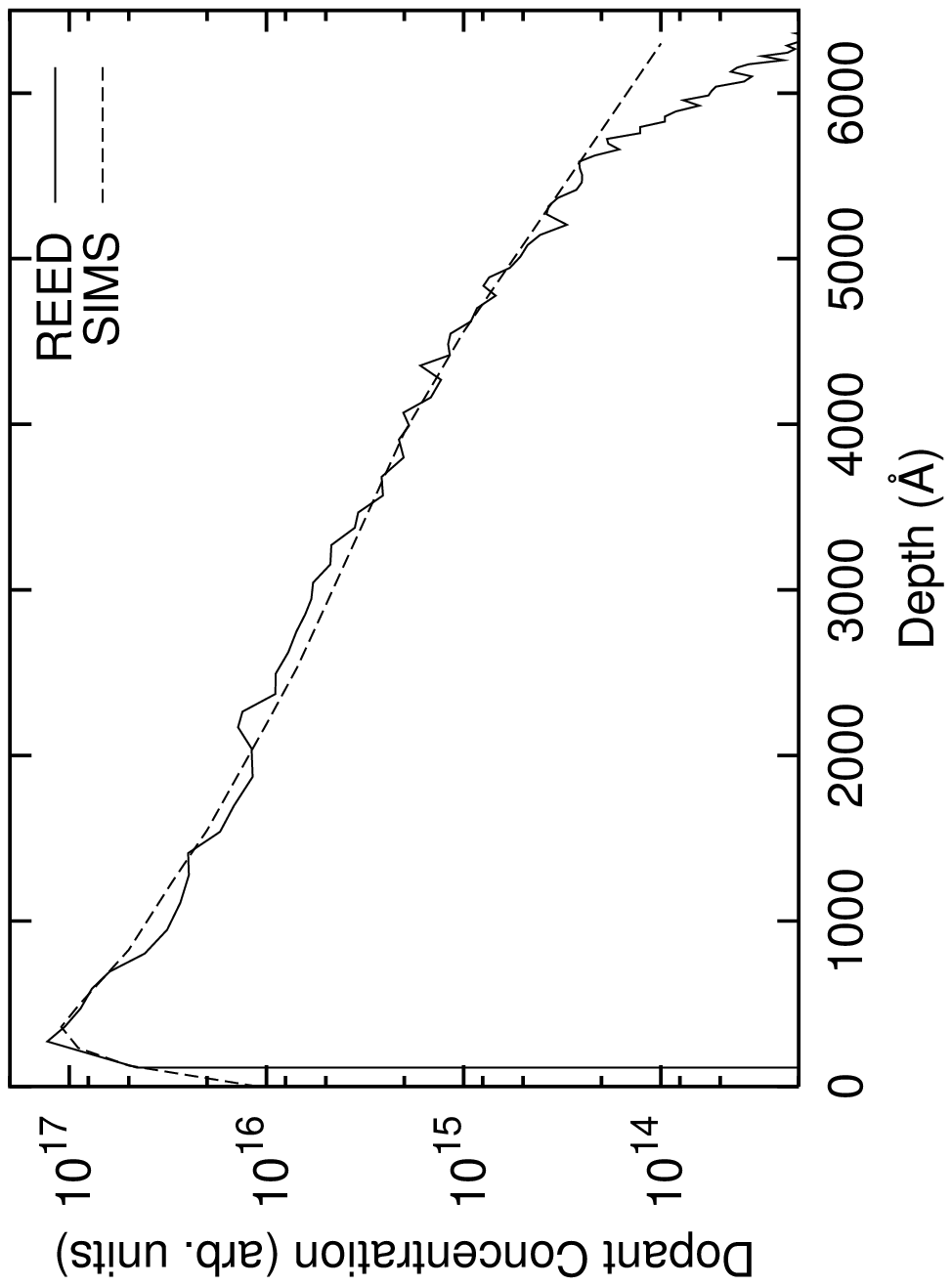}
\vspace{5.20 in}}

\caption{
The calculated and experimental\protect\cite{tas97} dopant profiles due to
50 keV As (0,0) into Si\{100\}.}
\label{As0t0r50K}

\end{figure}

\begin{figure}
\vbox{\vspace{-2.7 in} \hspace{0.5 in}
\includegraphics{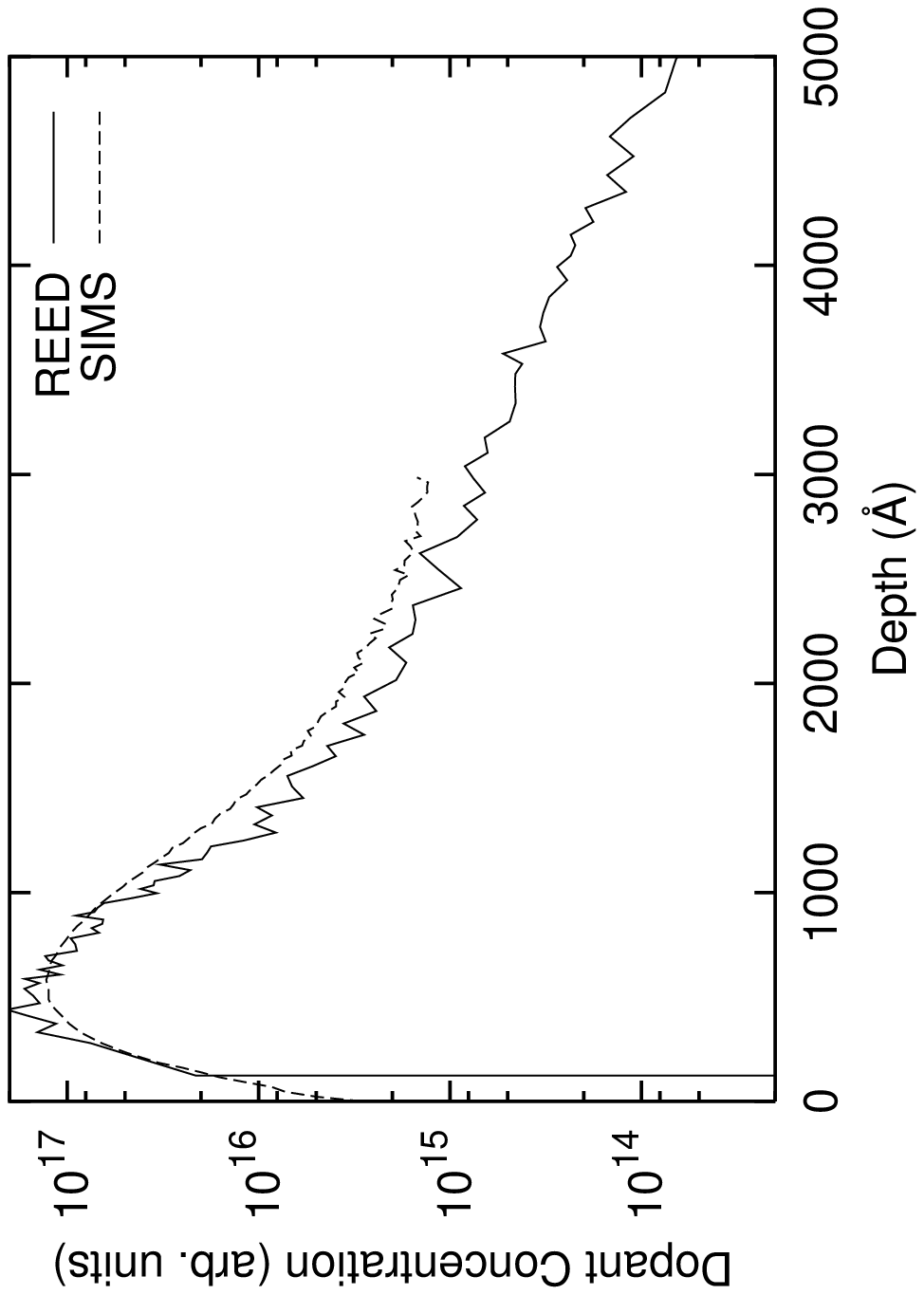}
\vspace{5.20 in}}

\caption{
The calculated and experimental\protect\cite{tas97} dopant profiles due to
100 keV As (8,30) into Si\{100\}.}
\label{As8t30r100K}

\end{figure}

\begin{figure}
\vbox{\vspace{-2.7 in} \hspace{0.5 in}
\includegraphics{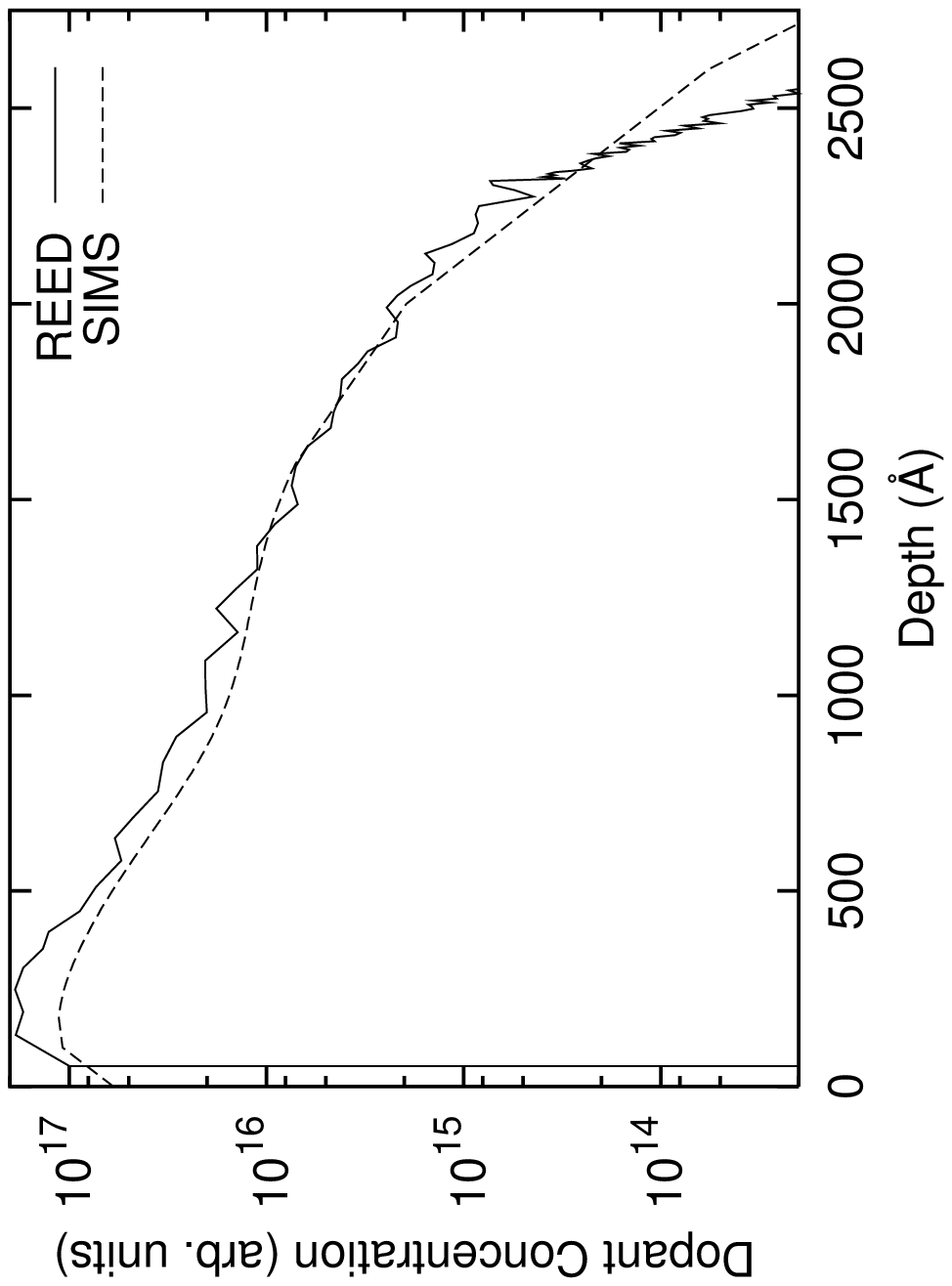}
\vspace{5.20 in}}

\caption{
The calculated and experimental\protect\cite{tas97} dopant profiles due to
15 keV P (0,0) into Si\{100\}.}
\label{P0t0r15K}

\end{figure}

\begin{figure}
\vbox{\vspace{-2.7 in} \hspace{0.5 in}
\includegraphics{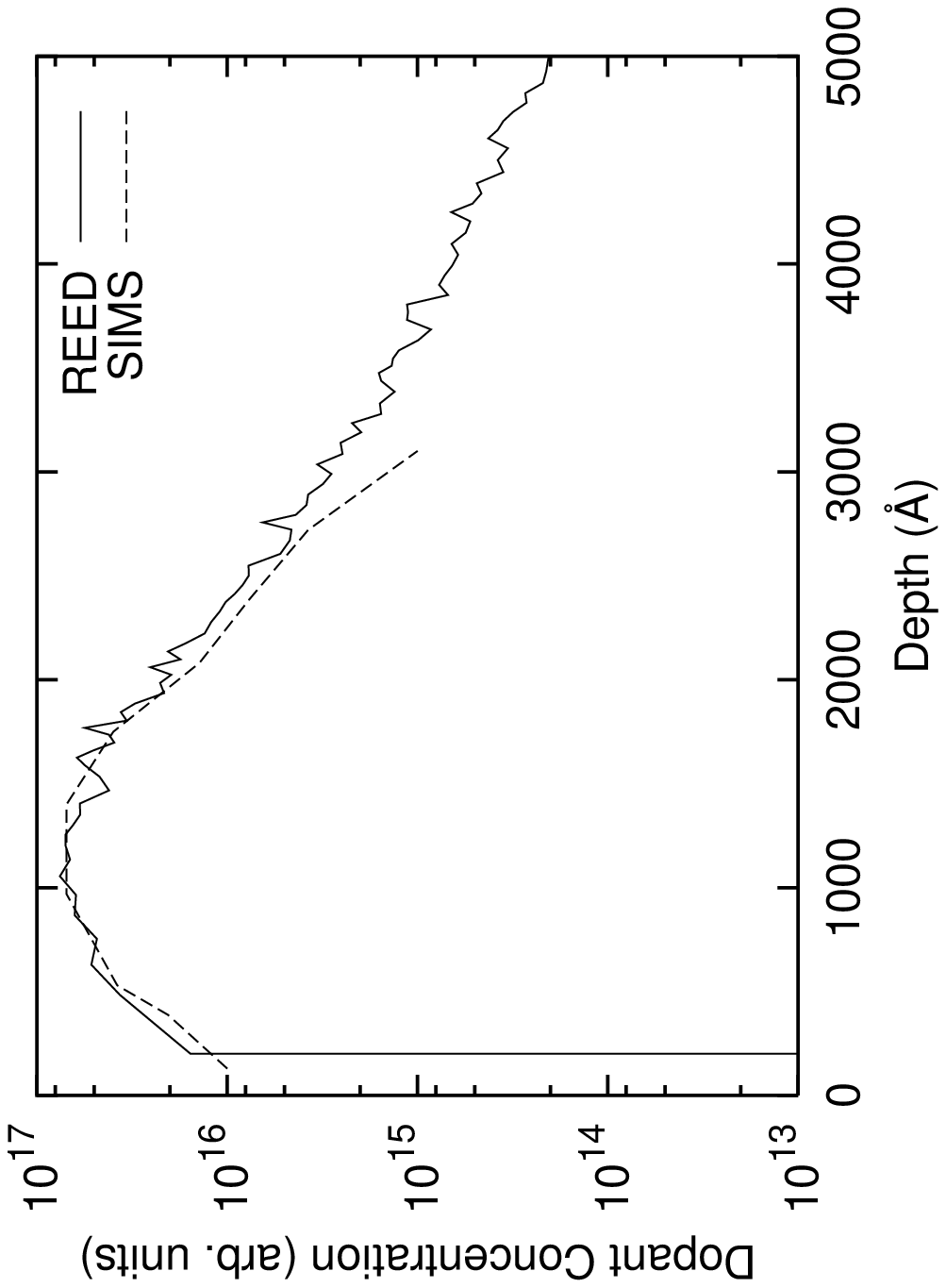}
\vspace{5.20 in}}

\caption{
The calculated and experimental\protect\cite{sch91} dopant profiles due to
100 keV P (10,15) into Si\{100\}.}
\label{P10t15r100K}

\end{figure}


\protect\pagebreak
\section{Tables}
%
%

\begin{table}
\caption{Rare event timing results for the case of 2 keV As
(7,0)\protect\tablenote{Run on a pentium pro., running Linux and g77.}.}
\label{rare}
\begin{tabular}{lrdrr}
$M$\tablenote{Split to a concentration $M$ orders of magnitude
less than the peak.}&\multicolumn{1}{c}{Run Time (s)}&
\multicolumn{1}{c}{Accuracy\tablenote{Number of orders of
magnitude before large uncertainty in the profile.}}&\multicolumn{1}{c}{Estimated time to 3 O.M.}&\multicolumn{1}{c}{Estimated time to 5 O.M.}\\ \tableline
0&10792&1&1079200&107920000\\
3&12136&3&12136&1213600\\
5&121777&5&-&121777\\
\end{tabular}
\end{table}

\begin{table}
\caption{Timing results for REED
simulations\protect\tablenote{Run on a pentium pro., running Linux and g77.}.}
\label{time}
\begin{tabular}{lddd}
Simulation\tablenote{All split to a concentration 5 orders of magnitude
less than the peak.}&\multicolumn{1}{c}{Run Time (s)}&
\multicolumn{1}{c}{Velocity ($\sqrt{{\text{keV}}/m_{\text{u}}}$)}&\multicolumn{1}{c}{Time/Velocity}\\ \tableline
2 keV As (7,0)&30650.28&0.23&133262.09\\
500 eV B (0,0)&17447.47&0.30&58158.23\\
15 keV P (0,0)&56544.42&0.70&81287.73\\
5 keV B (0,0)&61464.83&0.95&64699.82\\
5 keV B (10,22)&146762.03&0.95&154486.35\\
20 keV Al (0,0)&125437.13&1.22&102817.32\\
20 keV Al (45,45)&171572.93&1.22&140633.55\\
\end{tabular}
\end{table}

\end{document}